\documentclass[12pt]{article}
\usepackage{epsfig, graphics}
\usepackage{subcaption}

\usepackage{caption}

\usepackage{amsfonts}
\usepackage{eurosym}
\usepackage{mathtools, bm}
\usepackage{amssymb, bm}
\usepackage[latin1]{inputenc}

\def\bkR{{\rm I\kern-.17em R}}
\def\QED{\begin{flushright} QED \end{flushright}}

\def \1n{1\hskip -3pt \mbox{N}}

\def \Frac {\displaystyle \frac }
\def \Int {\displaystyle \int }
\def \Sum {\displaystyle \sum }

\newfont{\bbf}{cmbx12 scaled 1435}

\begin{document}
\thispagestyle{empty}

\vspace*{1.5cm}
\begin{center}

\setlength{\baselineskip}{.32in}
{\bbf Temporally Local Maximum Likelihood with Application to SIS Model}\\

\setlength{\baselineskip}{.26in}
\vspace{0.5in}

C. GOURIEROUX \footnote[1]{University of Toronto, Canada, Toulouse School of Economics (TSE) and CREST \\ {\it e-mail}:
{\tt gourierou@ensae.ca}.}, and J. JASIAK  \footnote[1]{York University, Canada, {\it e-mail}:
{\tt jasiakj@yorku.ca}.\\
The first author acknowledges financial support from the ACPR chair "Regulation and Systemic Risk", the ERC DYSMOIA and the Agence Nationale de la Recherche (ANR-COVID) [Grant ANR-17-EURE-0010].
The second author gratefully acknowledges financial support of the Natural Sciences and Engineering Council of Canada (NSERC).}\\
\vspace{0.1in}
\vspace{0.5in}
  
first version: January 2021\\
revised: \today\\

\vspace{0.1in}
\begin{minipage}[t]{12cm}
\small

The parametric estimators applied by rolling are commonly used in the analysis of time series with nonlinear features, such as structural change due to time varying parameters and local trends. This paper examines the properties of rolling estimators in the class of Temporally Local Maximum Likelihood (TLML) estimators.
We study the TLML estimators of constant parameters, stochastic and stationary parameters and parameters with the Ultra Long Run (ULR) dynamics bridging the gap between the constant and stochastic parameters. Moreover, we explore the properties of TLML estimators in an application 
to the Susceptible-Infected-Susceptible (SIS) epidemiological model and illustrate their finite sample performance in a simulation study.
\end{minipage}

\end{center}
\medskip
\textbf{Keywords~:} Local Maximum Likelihood, Rolling Estimator, Omitted Heterogeneity, Bias Reduction,  Generalized Linear Model, SIS Model, Logistic Growth, Ultra Long Run.

\newpage

\section{Introduction}
\setcounter{equation}{0}\def\theequation{1.\arabic{equation}}

The parametric estimators applied by rolling are commonly used for the analysis of time series with nonlinear features, such as structural change due to time varying parameters and local trends due to growth episodes, for example. The rolling approach is expected to adjust the estimates and predictors for changes in the trajectory of a process, while relying on simple computation and updating formulas.

The aim of this paper is to study  the properties of rolling estimators in the class of Temporally Local Maximum Likelihood (TLML) estimators [see e.g. Nicholls, Quinn (1980), Hastie, Tibshirani (1993), Cai (2007) for TLML in linear regression models and Anderes, Stein (2011) for random fields]. We study the TLML estimators of constant parameters, stochastic and stationary parameters and parameters with the Ultra Long Run (ULR) dynamics bridging the gap between the constant and stochastic parameters.

The TLML estimators are characterized by the underlying dynamic model and the selected sequence of weights. Our model of interest is an epidemiological SIS model defined by a set of deterministic differential equations. We consider its  discrete time stochastic version based on Markov chains   with heterogenous transition probability and the geometric and hyperbolic weights. We study the Poisson and Poisson-Gaussian approximations of the model that lead to generalized linear models (GLIM) and facilitate the numerical implementation of the TLML approaches.

In general, the functional TLML estimator indexed by time $t$ is close to a stationary process. The limiting stationary process can degenerate to a constant process if the weights are "local", i.e.  are sufficiently discounting observations far apart from time $t$. We derive the condition on the weights under which this degeneracy occurs and discuss the interpretations of the limiting constant, i.e. the pseudo-true value of the time varying parameter. Then, under the no-degeneracy condition on the weights, we discuss the asymptotic normality of TLML estimators. We also consider the case of stochastic parameters, which are smoothly varying and follow an Ultra Long Run process, in order to bridge the gap between models with constant parameters and models with stationary stochastically time varying parameters.

The paper is organized as follows.
The dynamic model with stochastic time varying parameters and the associated TLML estimators are introduced in Section 2. Section 3 describes the asymptotic properties of the TLML estimator under the joint stationarity assumption on the observations and stochastic parameters. We explain how these asymptotic results can be used in practice to build time varying confidence intervals for the TLML estimator.
An illustration based on the Susceptible-Infected-Susceptible (SIS) epidemiological model is provided in Section 4. This Section highlights the performance of the proposed approach in an epidemiological model.
Section 5 concludes. Appendix 1 provides the second-order asymptotic properties of the TLML estimator. It shows how  to get a more accurate approximation of the distribution of the estimator, by adjusting the TLML estimator for bias. Appendix 2 verifies the stationarity of the discrete time stochastic SIS model. Appendix 3 provides additional information on the quasi-collinearity and on the dynamic properties of the estimation errors.

\section{The Dynamic Model and the TLML Estimator}
\setcounter{equation}{0}\def\theequation{2.\arabic{equation}}
\subsection{The model}

The model involves $J$-dimensional observations $y_t, t=1,\ldots, T,$ and $K$-dimensional unobserved possibly time-varying parameters $\theta_t, t=1, \ldots,T.$ The model is defined by the conditional distribution of $y_t$ given $\underline{y_{t-1}} = (y_{t-1}, y_{t-2}, \ldots), \underline{\theta} = (\theta_t, t$ varying). The associated conditional density is specified as~:

\begin{equation}
  l(y_t | \underline{y_{t-1}} ; \underline{\theta} ) = l(y_t | y_{t-1}; \theta_t).
\end{equation}

\noindent It depends on the time varying parameter. More specifically, the above conditional density depends on the value of the parameter at time $t$ only. For expository purpose, we assume that process $(y_t)$ is a Markov \footnote{Alternatively, the right-hand side of (2.1) can be interpreted as the  conditional composite likelihood at order 1 [see Varin, Reid, Firth (2011)].} process given $\underline{\theta}$, with a time heterogenous transition probability.

We do not specify explicitly the dynamics of parameter $\theta_t$. Our statistical analysis assumes that:

i) $(\theta_t)$ is stochastic and strictly stationary;

ii) there is no parametric specification of the dynamic of $(\theta_t)$

\noindent Therefore, our approach differs from other types of nonparametric analysis which assume that $(\theta_t)$ is a smooth deterministic function of time [see e.g. Fan, Gijbels (1996), Cai et. al. (2000), Zhou, Wu (2010)] and from approaches that use specific state-space models, either deterministic  [see e.g. Cakmakli, Simsek (2020) for an application to epidemiology], or stochastic [see Kim, Nelson (1999), Canova, Perez, Forero (2015)].

\subsection{The TLML estimator}

Let us introduce a sequence of weights $w(h), h=0,1, \ldots,$ assumed nonnegative. The TLML estimator of $\theta$ is defined as~:

\begin{eqnarray}
  \hat{\theta}_T (w) & = & \arg \max_\theta \Sum^T_{t=1} \{ w(T-t) \log l(y_t |y_{t-1}; \theta)\} \\
  &=& \arg \max_\theta \Sum^T_{t=0} \{ w(h) \log l(y_{T-h} |y_{T-h-1}; \theta)\} \\
  &=&\arg \max_\theta \Sum^T_{h=0} \{ w(h) \log l(y_{T-1}|y_{T-h-1}; \theta)\}/ \Sum^{T-1}_{h=0} w(h).
\end{eqnarray}

\noindent The sequence $( \hat{\theta}_T (w)), T=1,2,....$ defines a nonparametric functional approximation of the stochastic process $(\theta_T, T=1,2,...)$. As $\theta_T$ is random, $\hat{\theta}_T $ is a predictor of $\theta_T$. Nevertheless, we follow the common practice of referring to $\hat{\theta}_T $ as an "estimator" of $\theta_T$, even though $\theta_T$ is not deterministic.

We assume below that the solution $\hat{\theta}_T (w)$ exists and is unique \footnote{In a multivariate structural vector autoregressive model (VAR) without parameter heterogeneity, the parameter $\theta$ is not identifiable, although the dynamic of $(\theta_t)$ can become identifiable, if $\theta_t$ is stochastic [see Primiceri (2005)]. }.

\noindent The TLML estimator is a pseudo maximum likelihood estimator because:

i) the heterogeneity of $\theta$ is omitted.

ii) the dated log-likelihoods are weighted, so that higher weights are assigned to the most recent current and lagged observations whenever the sequence $w(h)$ is decreasing in $h$. As we consider a dynamic framework for tracing the evolution of latent stochastic parameter in real time, a one-sided kernel is selected \footnote{Our analysis differs from the cases where the parameters vary with an exogenous variable and a continuum of values of this variable are asymptotically observed.}. The analysis in real time differs from the ex-post analysis, in which a long period of time is considered to detect either a small number of switching regimes with either sudden or smooth transitions [see, e.g. Francq, Gautier (2004)], or recurrent events in cross-sectional data [see, Yu et. al. (2013)], or to develop tests of constant parameters [Fan, Zhang (2000)].
Whereas a large part of the literature assumes longitudinal data, where many subjects are observed at multiple times [see, Fan, Zhang (2000), Lin, Ying (2001)], {\bf we consider a pure time series framework with only one available realization path}.

\noindent The following Sections 3 and 4 examine the impact of this double misspecification.

\subsection{The weights}

Let us introduce the cumulated weights and cumulated square weights~:

\begin{equation}
  W_T = \Sum^{T-1}_{h=0} w(h),\;\;\; W^{(2)}_T = \Sum^{T-1}_{h=0} [w(h)]^2.
\end{equation}

\noindent Various weighting schemes can be considered:\vspace{1em}

\textbf{Example 1~: Unweighted estimator}

\noindent This case corresponds to a standard ML estimator with omitted heterogeneity. We have : $w(h) = 1, \forall h, W_T = W^{(2)}_T = T.$\vspace{1em}

\textbf{Example 2~: Rolling weighted estimator}

\noindent This case arises when the weights are zero for $h$ sufficiently large: $w(h) = 0$, if $ h\geq H$. Then, $W_T = W_H, W^{(2)}_T = W^{(2)}_H$, if $T \geq H$.\vspace{1em}

\textbf{Example 3~: Geometric weights}

\noindent Let us assume : $w (h) = \rho^h$, with $0<\rho <1.$ We have~:

$$
W_T = \Sum^{T-1}_{h=0} \rho^h = \Frac{1-\rho^T}{1-\rho}, W^{(2)}_T = \Sum^{T-1}_{h=0}  \rho^{2h} = \Frac{1-\rho^{2T}}{1-\rho^2}.
$$

\textbf{Example 4~: Hyperbolic weights}

\noindent If $w(h) = (1/h)^c, c >0$, we see that~:

$\lim_{T\rightarrow \infty} W_T$ exists if $c>1$; otherwise, we have~: $W_T = 0 (\log T),$ if $c=1$,

$$
W_T = 0 (T^{1-c}), \mbox{if}\: 0 \leq c <1.
$$

\noindent In the following Sections 3.1, 3.2, we consider a fixed weighting function $w(.)$, which does not depend on the number of observations $T$. In epidemiological studies, the rolling is often performed over a window of one week, i.e. $H=7$ days [se e.g. Shapiro et. al (2020), PHO (2021), p.13, for COVID-19 analysis]. Later on, in Sections 3.3, 3.4, we introduce a kernel which allows for a well-chosen dependence on the number of observations. 

\section{Asymptotic Properties of the TLML Estimator Under Stationarity}

The asymptotic properties of the TLML are derived in this Section under the following stationarity assumption~:\vspace{1em}

\noindent \textbf{Assumption A.1 : Joint Stationarity} The joint process $(y_t, \theta_t)$ is strictly stationary.\vspace{1em}

Assumption A.1 includes the special case of $\theta_t = \theta_\infty$  independent of $t$ and a strictly stationary process $(y_t)$. In this special case, there is no omitted heterogeneity in the TLML approach and the likelihood is well-specified. Generally, Assumption A.1 allows for other dynamics of $\theta_t$, such as : i) i.i.d. stochastic parameters [Nicholls, Quinn (1980)], ii) Gaussian AR(1) stochastic parameter $\theta_t = \mu + \rho (\theta_{t-1} - \mu) + \sigma u_t$, where $u_t$ are i.i.d. N(0,1), and iii) ultra long run (ULR) process $\theta_{t,T} = \mu + \rho_T (\theta_{t-1,T} - \mu) + \sigma_T u_{tT}$, where $\rho_T$ tends to 1 and $\sigma_T$ tends to 0 at appropriate speeds when $T$ tends to infinity. The ULR process accommodates small smoothed stationary deviations of the parameter from a constant path [see Gourieroux, Jasiak (2021) for ULR processes, Froeb, Koyak (1994) for the notions of smoothness in stochastic time series].

Assumption A.1 considers  stationary stochastic parameters. As mentioned earlier, it excludes the parameters defined as a smooth deterministic function of time, considered in the literature on nonparametric estimators of time varying coefficients [see e.g. Cai (2007), Zhou, Wu (2010)] that introduces nonstationary features and disregards the uncertainty on the parameter in the long run.

Below,  we derive the main asymptotic results, which are the conditions of consistency, i.e. convergence to a pseudo-true value, the asymptotic normality of the local estimator under consistency [see  Appendix 1 for the higher order expansion]. We also discuss the stationarity of the sequence of local estimators when the consistency condition is not satisfied. We study the ULR process of $\theta_t$ to bridge the gap between the constant parameters and stationary stochastic parameters. We restrict our attention to regularity conditions, which are necessary for clarity of the results.

\setcounter{equation}{0}\def\theequation{3.\arabic{equation}}

\subsection{Consistency}

Suppose, the weights are fixed and do not depend on the number of observations. The consistency of the estimator is deduced from the asymptotic behaviour of the optimization criterion under a suitable normalization.

\noindent  Let us introduce the weighted criterion function:

\begin{equation}
  L_T (w;\theta) = \Sum^{T-1}_{h=0} \{ w(h) \log l(y_{T-h} | y_{T-h-1}; \theta)\}/\Sum^{T-1}_{h=0} w(h).
\end{equation}

\noindent Next, we assume that the moments of $\log l(y_t |y_{t-1};\theta)$ exist up to order 2. Then, the first-order moment is

\begin{equation}
  E_0 L_T (w;\theta) = E_0 \log l(y_t|y_{t-1};\theta),
\end{equation}

\noindent and the second-order moment is:

\begin{equation}
  V_0 L_T (w;\theta) = \Frac{\Sum^{T-1}_{h=0} \Sum^{T-1}_{k=0} [w (h) w(k) \gamma (h-k;\theta)]}{[\Sum^{T-1}_0 w(h)]^2},
\end{equation}

\noindent where $\gamma (h;\theta) = Cov_0 [\log l(y_t|y_{t-1};\theta), \log l(y_{t-h} | y_{t-h-1}; \theta)].$  and $E_0, V_0$ are the expectation and variance computed from the true distribution of process $(y_t)$.  This true distribution involves both the true conditional transition (2.1) and the true stationary distribution of $\theta_t$.\vspace{1em}

\noindent Let us assume that a geometric mixing condition holds, for ease of exposition.\vspace{1em}

\textbf{Assumption A.2~: Geometric mixing}

The process $(y_t, \theta_t)$ is geometrically mixing with geometric order $r$.\vspace{1em}

\noindent  Then, we have~:

$$
V_0 L_T (w;\theta) \leq \gamma (0;\theta) \Sum^{T-1}_{h=0} \Sum^{T-1}_{k=0} [w(h) w(k) r^{|h-k|}]/ [\Sum^{T-1}_{h=0} w(h)]^2,
$$

\noindent and

\begin{eqnarray}
  V_0 L_T (w;\theta) &\leq& \gamma (0;\theta) W^{(2)}_T (r) / (W_T)^2, \\
  \mbox{with}\; W^{(2)}_T (r) & = & \Sum^{T-1}_{h=0} \Sum^{T-1}_{k=0} (w(h) w(k) r^{|h-k|}).
\end{eqnarray}

\noindent  We deduce the following result:\vspace{1em}

\textbf{Proposition 1~:} If $W^{(2)}_T (r)/(W_T)^2$ tends to zero when $T$ tends to infinity, then the finite sample objective function $L_T (w;\theta)$ tends to the limiting objective function $L_\infty (w;\theta) = E_0 \log l(y_t|y_{t-1}; \theta)$, where $E_0$ denotes the expectation taken with respect to the true joint stationary distribution of $(y_t, y_{t-1})$, which depends on the true underlying dynamics of the stochastic parameter.\vspace{1em}

In particular this limiting function does not depend on the sequence of weights. Then, by applying the standard Jennrich's argument [Jennrich (1969), Andrews (1987)], the solution of the finite sample optimization problem will tend to the solution of the asymptotic problem.\vspace{1em}

\textbf{Corollary 1~:} If $W^{(2)}_T (r)/ (W_T)^2$ tends to zero, when $T$ tends to infinity, then $\hat{\theta}_T (w)$ (exists asymptotically and) is consistent of \\
 $\theta^*_\infty = \arg \max_\theta E_0 \log l(y_t | y_{t-1}; \theta).$\vspace{1em}

\textbf{Remark 1~:} If the joint process $(y_t, \theta_t)$ is a sequence of i.i.d. variables, we have $W^{(2)}_T (r) = W^{(2)}_T$, and the sufficient condition for consistency to $\theta^*_\infty$ becomes $W^{(2)}_T / (W_T)^2$ approaching zero when $T$ tends to infinity.\vspace{1em}

\textbf{Remark 2~:} If there is no parameter heterogeneity $\theta_t = \theta_\infty, \forall t,$ then $\theta^*_\infty = \theta_\infty$, by the property of the Kullback-Leibler information criterion. In the presence of heterogeneity, there is no notion of a true value of parameter $\theta$ and $\theta^*_\infty$ is a pseudo true value of this parameter that depends on the joint distribution of process $(y_t, \theta_t)$. In general, this pseudo-true value depends on both the true distribution of $(\theta_t)$ and the true transition (2.1). In particular $\theta^*_\infty$ is not equal, or even close, to the expected parameter value $E_0 \theta_t$.\vspace{1em}

The condition on the weights given in Proposition 1 is easy to interpret. The TLML estimator is convergent if  the observations far from $T$ are sufficiently down-weighted. More precisely, for $T$ large, the estimator $\hat{\theta}_T (w)$ is close to the virtual estimator~:

\begin{equation}
  \theta^*_T (w) = \arg \max_\theta \Sum^\infty_{h=0} \{ w (h) \log l(y_T | y_{T-h} ; \theta)\},
\end{equation}

\noindent computed with an infinite sum. This virtual estimator is a fixed function of the stationary process $(y_t)$. Therefore it is also stationary. More precisely, the joint process $(\theta_t^*(w), y_t, \theta_t)$ is strictly stationary. \vspace{1em}

\textbf{Proposition 2~:} Under the stationarity assumption A.1, the TLML estimator $\hat{\theta}_T(w)$ is equivalent to the virtual TLML estimator $\theta^*_T (w)$. This virtual functional estimator $(\theta^*_T (w), T$ varying) is such  that $(\theta_T^*(w), y_T, \theta_T)$ is a strictly stationary process.\vspace{1em}

\noindent The following two extreme cases can be distinguished: The process $[\theta^*_T (w)]$ is either:

i)  constant; Proposition 1  provides a sufficient condition for this property to hold and shows that $\theta^*_T (w) = \theta^*_\infty$.

ii) or stationary, but does not degenerate to a constant.\vspace{1em}

\noindent In the second case ii), the joint process $[\theta_T, \theta^*_T (w)]$ is also stationary, but $[\theta_T - \theta_T^*(w)]$ does not have mean zero, in general.\vspace{1em}

\textbf{Example 5 : Gaussian observations with mean heterogeneity}.\vspace{1em}

To illustrate the above results, let us consider the model with the observations that are i.i.d. with distribution $N (\theta_t, 1)$  conditional on $\underline{\theta} = (\theta_t)$. The TLML estimator is a weighted average of the observations~:

$$
\hat{\theta}_T (w) = \Sum^{T-1}_{h=0} [w(h) y_{T-h}]/ [\Sum^{T-1}_{h=0} w(h)].
$$

\noindent It can be written as~:

$$
\hat{\theta}_T (w) = \Sum^{T-1}_{h=0} (w(h) \theta_{T-h}) / [\Sum^{T-1}_{h=0} w(h)] + \Sum^{T-1}_{h=0} (w(h) u_{T-h}) / [\Sum^{T-1}_{h=0} w(h)],
$$

\noindent where $u_t = y_t - \theta_t$ are i.i.d. $N(0,1)$. Therefore, conditional on $\underline{\theta}$, the distribution of $\hat{\theta}_T (w)$ is normal with (conditional) mean $m_T (w; \underline{\theta}) = \Sum^{T-1}_{h=0} [w(h) \theta_{T-h}] / [\Sum^{T-1}_{h=0} w(h)]$ and variance $\sigma^2_T (w) = W^{(2)}_T / W^2_T$.\vspace{1em}

\noindent  The unconditional mean of the estimator is equal to~:

$$
E_0 \hat{\theta}_T (w) = E_0 m_T (w;\underline{\theta}) = E_0 \theta_T.
$$

\noindent  Therefore, the difference between the weighted local estimate and the time varying parameter has mean zero. That implies that the functional predictor $(\hat{\theta}_T)$ is unbiased of $(\theta_T)$.

i) Rolling weighted estimator (see Example 2).\vspace{1em}

For $T \geq H$, we have~: $\hat{\theta}_T (w) = \Sum^{H-1}_{h=0} [w(h) y_{T-h}]/[\Sum^{H-1}_{h=0} w(h)].$

\noindent It follows that $(\hat{\theta}_T (w))$ is a moving-average (MA) transformation of process $(y_T)$ with a finite moving average order equal to $H$. In particular, it does not converge when $T$ tends to infinity.\vspace{1em}

ii) Geometric weights (Example 3).\vspace{1em}

\noindent  We get :

$$
\hat{\theta}_T (w) \simeq \theta^*_T (w) = \Sum^\infty_{h=0} (\rho^h y_{T-h}) (1-\rho),
$$

\noindent which is a nondegenerate moving average MA($\infty$) transformation of an infinite MA order. The weights satisfy:

$$
W^{(2)}_T/(W_T)^2 = \Frac{1-\rho^{2T}}{(1-\rho^T)^2} \Frac{(1-\rho)^2}{1-\rho^2},
$$

\noindent which tends to $\lim_{T\rightarrow \infty} W^{(2)}_T/(W_T)^2 = \Frac{1-\rho}{1+\rho} \neq 0$, if $ 0 \leq \rho < 1.$\vspace{1em}

iii) Hyperbolic rates (Example 4)\vspace{1em}

\noindent The consistency, i.e. the convergence to a pseudo true value, can be reached with a hyperbolic weight $w(h) = (1/h)^c$ . Indeed, we observe the following asymptotic behaviour~:

$$
\begin{array}{lcll}
  W^{(2)}_T / (W_T)^2 & = & 0 (T^{1-2c}/T^{2-2c}) = 0 (1/T) = o(1),& \mbox{if}\; c <1/2, \\ \\
                      & = & 0 (\log T/T) = o(1), \; &\mbox{if}\; c <1/2, \\ \\
                      & = & 0 (1/T^{2-2c}) = o(1), \;& \mbox{if}\; \Frac{1}{2} < c <1,  \\ \\
                      & = & 0 (1/\log T) = o(1), \; &\mbox{if}\; c =1,\\ \\
                      & = & 0 (1), \; &\mbox{if}\; c >1.
\end{array}
$$

\noindent  The consistency is achieved when $c\leq1,$ with $\lim_{T\rightarrow \infty} \hat{\theta}_T (w) = E_0 \theta_t$.\vspace{1em}

\noindent  In some sense, if $c \leq 1$, the TLML estimator is not sufficiently local, as it tends to a global summary of the joint distribution of $(y_t, \theta_t)'s$, $\forall t$. Although the estimator does not converge when it is "too local", it has a non-degenerate distribution that can be used for statistical inference.

\subsection{Asymptotic Normality}

When the TLML estimator converges to a pseudo-true value $\theta^*_\infty$, we can write the first-order expansion of the first-order conditions (FOC) of the objective function. The FOC are~:

\begin{equation}
  \Sum^{T-1}_{h=0} \{ w (h) \Frac{\partial \log l}{\partial \theta} [y_{T-h} | y_{T-h-1}; \hat{\theta}_T (w)]\} = 0.
\end{equation}

\noindent The FOC can be expanded in a neighbourhood of the pseudo-true value $\theta^*_\infty$:

\begin{eqnarray}
  && \Sum^{T-1}_{h=0} \{ w(h) \Frac{\partial \log l}{\partial \theta} [y_{T-h} | y_{T-h-1}; \theta^*_\infty] \} \nonumber \\
  &+& \Sum^{T-1}_{h=0} \{ w(h) \Frac{\partial^2 \log l}{\partial \theta \partial \theta'} [y_{T-h} | y_{T-h-1}; \theta^*_\infty]\} (\hat{\theta}_T (w) - \theta^*_\infty) = o_P,
\end{eqnarray}

\noindent where $o_P$ is negligible in probability with respect to the components of the left hand side of the equation.\vspace{1em}

\noindent  Let us introduce the following matrices and vectors:

$$
\begin{array}{lcl}
J (\theta^*_\infty) & = & E_0 \left[ - \Frac{\partial^2 \log l}{\partial \theta \partial \theta'} (y_t | y_{t-1}; \theta^*_\infty) \right],\\ \\
I(h;\theta^*_\infty) & = & cov_0 \left[ \Frac{\partial \log l}{\partial \theta} (y_t | y_{t-1} ; \theta^*_\infty), \Frac{\partial \log l}{\partial \theta'} (y_{t-h}, y_{t-h-1}; \theta^*_\infty)\right], \\ \\
X_T &=& \left[ \Sum^{T-1}_{h=0} \Sum^{T-1}_{k=0} [w(h) w(k) I (h-k;\theta^*_\infty)]\right]^{-1/2} \Sum^{T-1}_{h=0} \left( w(h) \Frac{\partial \log l}{\partial \theta} (y_{T-h} | y_{T-h-1}; \theta^*_\infty) \right). \\ \\
&\equiv & \left[ I_T (w;\theta^*_\infty)\right]^{-1/2} \Sum^{T-1}_{h=0} \left[ w(h) \Frac{\partial \log l}{\partial \theta} [y_{T-h} | y_{T-h-1}; \theta^*_\infty)\right],
\end{array}
$$

\noindent where $J (\theta^*_\infty)$ and $ I(h;\theta^*_\infty)$ are independent of both time and weights.
Therefore the expansion (3.8) can be rewritten as~:

\begin{equation}
  X_T = [I_T (w;\theta^*_\infty)]^{-1/2} J (\theta^*_\infty) W_T (\hat{\theta}_T (w) - \theta^*_\infty) + o_P,
\end{equation}

\noindent where $X_T$ is asymptotically $N(0, Id)$.\vspace{1em}

\textbf{Proposition 3~:} Under Assumptions A-1, A-2, we have~:

$$
[I_T (w;\theta^*_\infty)]^{-1/2} J (\theta^*_\infty) W_T (\hat{\theta}^{(w)}_T - \theta^*_\infty) \approx N (0, Id).
$$

\vspace{1em}

\noindent This is the case of double misspecification, which results from estimating the parameter  
as if the parameter were constant and as if the weighted log-likelihood function with the constant parameter were
well-specified.
In general, this double misspecification affects the limit of the estimator, its speed of convergence  and entails the sandwich form of its asymptotic variance-covariance matrix [Huber (1967), White (1982)]. Let us discuss these effects in more detail.

Without the time variation of the parameter $\theta_t = \theta_\infty, \forall t, \theta^*_\infty = \theta_\infty$ and $\left[\Frac{\partial \log l}{\partial \theta} (y_t | y_{t-1}; \theta_\infty)\right]$ is a martingale difference sequence (MDS) and the FOC are based on martingale estimating equations, using the terminology of Godambe, Heyde (1987). The mixing coefficient of this sequence corresponds to $r=0$ and the condition for convergence is $\lim_{T\rightarrow \infty} W^2_T / W^{(2)}_T = 0.$

This MDS condition implies also that $I (h;\theta^*_\infty) = I (h;\theta_\infty) = 0,$ if $h \neq 0$. Moreover we have $I (0; \theta_\infty) = J (\theta_\infty)$. We deduce the following Corollary:\vspace{1em}

\textbf{Corollary 2~:} Let us assume A.1 and the absence of heterogeneity $(\theta_t = \theta_\infty, \forall t),$ then, if $\lim_{T\rightarrow \infty} W^2_T / W^{(2)}_T = 0$, we have~:

$$
J^{1/2}(\theta_\infty) \Frac{W_T}{\sqrt{W^{(2)}_T}} (\hat{\theta}_T (w) - \theta_\infty) \approx N(0, Id).
$$

\noindent Under the assumptions of Corollary 2, the presence of local weights has no effect on the limit of the estimator. It does not entail a "sandwich" formula of asymptotic variance, but can change the speed of convergence of the estimator : the more local the weights are, the slower the speed of convergence.\vspace{1em}

\textbf{Remark 3~:} The computation of the TLML estimator at date $T$ is easy, but can become costly if it has to be computed ex-post from a large number of dates. In that case, this estimator can be replaced by an Iterative Local Maximum Likelihood (ILML) estimator, which is obtained from a one-step of the Newton-Raphson maximization algorithm applied to the local log-likelihood $L_T (w;\theta)$, with  starting value $\hat{\theta}_{T-1} (w)$ [see e.g. Cai, Fan, Li (2000), Section 2.2].

\subsection{The Frontier Between Local and Global Analysis}

The asymptotic distribution of Proposition 3 is valid when the weights are sufficiently global. Otherwise, it follows from Proposition 2 that the sequence of TLML estimators is stationary, but the distributional properties of this sequence cannot be derived analytically for a time varying stationary stochastic parameter. These distributional properties can only be explored in a Monte-Carlo experiment by using different scenarios for the dynamics of $\theta$ (see Section 4 for a simulation study).

It is, however, possible to examine analytically what may arise at the "frontier" between the local and global approaches by considering ULR processes of $(\theta_t)$ [Gourieroux, Jasiak (2021)]. For expository purpose, we consider the example of Gaussian observations with mean heterogeneity (Example 5), where $y_t \sim N(\theta_t,1)$. We assume a stationary Gaussian ULR process of $(\theta_t)$ (a triangular array, more precisely):

\begin{equation}
  \theta_{t,T} = \mu + \rho_{t,T} (\theta_{t-1,T} - \mu) + \sigma_{t,T}\, v_{tT}, \;\;\; v_{tT} \sim \mbox{IIN} (0,1).
\end{equation}

\noindent If $\rho_{t,T}$ tends to $1, \sigma_{t,T}$ tends to 0 when $T$ tends to infinity, then the local-to-unity/small sigma autoregressive dynamic process tends to a time invariant trajectory $\theta_t =\theta_0, \forall t.$ The level $\theta_0$ of this trajectory is stochastic and differs from its theoretical mean $\mu = E_0 \theta_t = E_0 y_0$. Thus at the limit, we get a purely predictable process in the terminology of Wold decomposition. An appropriate choice of this  type of triangular array can be derived from a continuous time latent dynamic. Let us introduce a latent stationary Ornstein-Uhlenbeck process $(\tilde{\theta}(\tau))$ such that~:

\begin{equation}
  d\tilde{\theta} (\tau) = -k [\tilde{\theta} (\tau) - \mu] d\tau + \eta d\tilde{W} (\tau),
\end{equation}

\noindent where $[(\tilde{W} (\tau)$] is a Brownian motion, $k, \eta$ are positive parameters. The stationary distribution of $\tilde{\theta}$ is Gaussian with mean $\mu$ and variance $\eta^2/2k.$ Then, process $\theta_{t,T} =\tilde{\theta}(t/T)$ satisfies a discrete time stationary autoregressive dynamic (3.10) with $\rho_{t,T} = \exp (-k/T)$, $\sigma^2_{t,T} = \eta^2 \Frac{1-\exp (-2k)}{2k} \Frac{1-\exp (-2k/T)}{1-\exp (-2k)}$ and the same stationary distribution as $\tilde{\theta} (\tau)$. Thus $\rho_{t,T}$ tends to 1 and $\sigma^2_{t,T}$ tends to zero at speed $1/T$. When $T$ increases there is less stochastic time heterogeneity and the analysis with a given sequence of weights will become more global.\vspace{1em}

Let us now introduce a sequence of weights also indexed by $T$ as~:

\begin{eqnarray}
  w_T (h) & = & \tilde{w} (h/T), \;\mbox{for}\; h \leq \tilde{H}_T \sim c T, \\
      & = &0, \hspace{1,5cm}\mbox{otherwise,}  \nonumber
\end{eqnarray}

\noindent Hence, at the limit we get a purely predictable process.

When $T$ increases, the sequence of weights tends to a constant infinite sequence $\lim_{T\rightarrow \infty} w_T (h) = \tilde{w} (0).$\vspace{1em}

\noindent  Then, the TLML estimator is equal to~:

\begin{equation}
  \hat{\theta}_T (w_T) = \Frac{\Frac{1}{T} \Sum^{H_T}_{h=0} \left[\tilde{w} (\Frac{h}{T}) \tilde{\theta} (1-\Frac{h}{T})\right]}{\Frac{1}{T} \Sum^{H_T}_{h=0} \tilde{w} (\Frac{h}{T})} +
  \Frac{\Sum^{H_T}_{h=0} [\tilde{w} (\Frac{h}{T}) u_{T-h}]}{\Sum^{H_T}_{h=0} \tilde{w} (\Frac{h}{T})},
\end{equation}

\noindent where $H_T = \min (T-1, \tilde{H}_T) \sim \min (T-1, c T) = cT$ for large $T$ and the $u'_t s$ are i.i.d. standard normal, independent of process $\tilde{\theta}(.)$. We deduce the following asymptotic behaviour~:\vspace{1em}

\textbf{Proposition 4~:} In the Gaussian model with ULR mean process, the TLML estimator with time varying weights tends to~:

$$
\lim_{T \rightarrow \infty} \hat{\theta}_T (w_T) = \Frac{\Int^c_0 \tilde{w} (\tau) \tilde{\theta} (1-\tau) d\tau}{\Int^c_{0} \tilde{w} (\tau) d\tau} \equiv B(c).
$$

\textbf{Proof~:} The numerator and denominator of the first component in the RHS of (3.13) are Riemann sums that converge to their associated (stochastic) integrals [Hansen (1992)].
The second component is close to $\Frac{1}{T} \Sum^{cT}_{h=0} u_{T-h} \sim c E u_t = 0$, by the Law of Large Numbers.

\QED

\noindent As a consequence of weights change, the sequence of local estimators is no longer stationary. However, for large $T$, this stochastic sequence tends to a stochastic level that depends on the weights and the dynamic of $\tilde{\theta} (\tau)$, i.e. on the long run dynamic of $\theta_{t,T}$.\vspace{0.1em}

This example shows that the consistency result of Section 3.1 can be extended to a  stochastic parameter following a ULR process with accordingly chosen weights. Note that the choice of weights as in (3.12) supposes a large number of weighted observations.

The analytical derivation given above has to be used with caution when applied by rolling over a rather short window. For a fixed window of width $H$, say, and equal weights, the estimator is~:

$$
\begin{array}{lcl}
\hat{\theta}_T (w) & = & \Frac{1}{T} \Sum^{H-1}_{h=0} \tilde{\theta} (1-\Frac{h}{T}) + \Frac{1}{H} \Sum^{H-1}_{h=0} u_{T-h} \\ \\
&\sim & \theta_T + \Frac{1}{H} \Sum^{H-1}_{h=0} u_{T-h}.
\end{array}
$$

\noindent The short window modifies the distribution of $\hat{\theta}_T (w) - \theta_T$, which is now equal to $N(0,1/H).$\vspace{1em}

This Gaussian example is somehow misleading, when the associated weighted local maximum likelihood estimators are interpreted as weighted averages. Similar results could be derived for more complicated examples, when the pseudo first-order conditions are not linear with respect to the parameters. Then, the limiting distribution of $\hat{\theta}_T (w) - \theta_T$ will not have mean zero.

\subsection{Confidence Intervals}

Let us now explain how the asymptotic results derived in Sections 3.1-3.3 can be used to build time varying confidence intervals for the TLML estimators. We first consider a constant parameter and a stationary stochastic parameter models. Next, we show how to bridge the gap between these models. Scalar parameters are assumed for ease of exposition.

\subsubsection{Model with constant parameters}

The constant parameter model is a standard asymptotic framework. A confidence interval  (CI) for $\theta_{\infty}$ can be based on Corollary 2. It is computed as:

$$\left[ \hat{\theta}_T(w) \pm 2 \frac{\sqrt{W_T^{(2)}}}{W_T} \hat{J}_T^{-1/2} [ \hat{\theta}_T(w)] \right], $$

\noindent where

$$  \hat{J}_T (\theta) = - \sum_{h=0}^{T-1} w(h) \frac{d^2 log l}{d \theta^2} (y_{T-h} | y_{T-h-1}; \theta)/ \sum_{h=0}^{T-1} w(h).$$

In practice, the same CI are used for models with time-varying parameters. This approach is valid when $\theta_t$ does not vary too much in the neighbourhood of time $T$. It is not valid, otherwise. The practice of estimating an asymptotic CI can be partly improved by adjusting it for "finite sample" bias [see Appendix 1 for this adjustment to the weighted (pseudo) log-likelihood].

\subsubsection{Model with stationary stochastic parameter}

The stationary stochastic parameter model entails the curse of dimensionality, due to the unknown distribution of process $(\theta_t)$. Therefore, a reasonable CI cannot be provided. However, despite this identification issue, information on the accuracy of estimators can be revealed. Two approaches can be followed, which are described below and used in Section 4.  

\medskip

i) {\bf Prediction accuracy}

\noindent Instead of predicting the future values of parameter $\theta_T$, one can focus on the short-run prediction of $y$. For a constant parameter model, $y_{T+1}$ can be predicted by $m(y_T; \theta) = E(y_{T+1}|y_T; \theta)$. When $
\theta_t$ is time varying, $y_{T+1}$ can be predicted by $\hat{y}_{T+1} = m[y_T; \hat{\theta}_T(w)]$. The joint process $(y_{T+1}, \hat{y}_{T+1})$ is stationary by Proposition 2. Then we can estimate nonparametrically the conditional distribution of $( y_{T+1}, \hat{y}_{T+1})$ given $y_T$, or given $(y_T, \hat{y}_{T})$ and build a conditional prediction interval.

\medskip

ii) {\bf Scenarios}

If we focus on parameter $\theta_T$ only, we can consider different dynamic models for the stationary evolution of $(\theta_t)$. As the models are easy to simulate (see the application to SIS modelling), it is possible to draw the time-varying parameters $(\theta_t^s, \; t=1,...,T)$, compute $(y_t^s, \; t=1,...,T)$ corresponding to this draw
and then calculate the sequence of dynamic TLML estimators $[\hat{\theta}_T^s,\; t=1,...,T]$. Given that $[\theta_T, \hat{\theta}_T(w)], \; T$ varying, is stationary, we can find by averaging over a large number of replications the marginal distribution of $( \hat{\theta}_T(w) - \theta_T)$, for example [see Figures 10-11 in the application to SIS modelling]

\subsubsection{ULR dynamics for bridging the gap}

The ULR dynamics introduced in Section 3.3 bridges the gap between the model with constant parameters and analytical CI formula (Section 3.4.1), and the model with stationary stochastic parameters and the curse of dimensionality (Section 3.4.2).

Proposition 4 shows that, although the TLML estimator is not consistent, its distribution is equal to the distribution of a ratio of stochastic integrals, whose dynamics depend on the underlying parameters $k, \eta$ of the Ornstein-Uhlenbeck process (3.11). Proposition 4 allows however for different choices of $H_T \approx cT$, i.e. of parameter $c$. Let us compute TLML estimates corresponding to $K$ values $c_1,...,c_K: \hat{\theta}_T (c_k; w_T), \; k=1,...,K$.

For large $T$, the joint asymptotic distribution of the $K$-dimensional vector of TLML estimators is equal to the distribution of the vector of ratios of stochastic integrals $[B(c_1),...,B(c_K)]$, where $B(c)$ is defined in Proposition 4. 

This asymptotic distribution is unknown, as it depends on the two unknown parameters $k, \eta$. These parameters can be estimated by applying the maximum likelihood method to the observed $\hat{\theta}_T (c_k; w_T), k=1,...,K$ with the Ornstein-Uhlenbeck likelihood. Given these, one can estimate the limiting distribution 
 of $\hat{\theta}_T (w_T)$.
 
The estimators of parameters $k, \eta$ are not expected to be accurate, as they are based on a finite sample of
summary statistics  $\hat{\theta}_T (c_k; w_T), \; k=1,...,K$. The treatment of this "finite sample" issue is out of the scope of the present paper. However, the confidence intervals can be obtained by applying a test inversion bootstrap approach [Carpenter (1999)].

\section{An Illustration}
\setcounter{equation}{0}\def\theequation{4.\arabic{equation}}

To illustrate the finite sample and asymptotic properties of the TLML estimator, we consider below an epidemiological dynamic model with two compartments $S=1$, susceptible individuals, $I =2$, infected individuals, where after the infectious period the individual is not immune and becomes again susceptible. This type of model is known under the acronym SIS [Susceptible-Infected-Susceptible] [see e.g. Brauer et al. (2008) for a discussion of SIS models]. The SIS models are used in applications to some sexually transmitted diseases and bacterial diseases [Hethcote, Yorke (1994)], such as tuberculosis, meningitis and gonorrhea. We consider the SIS model in  our illustration due to its simplicity allowing us to explain its stochastic extensions and to perform statistical inference. This model is also interesting because the trajectories of the counts of infected have nonlinear dynamic features, such as peaks and cluster effects that may render inaccurate the TLML approaches during some episodes of an epidemic.

\subsection{The (Stochastic) SIS Model}

\subsubsection{The differential equation}

In the epidemiological literature the SIS model is usually defined in continuous time, assuming a population of infinite size and deterministic dynamics. Let $p_j (t)$  denote  the proportion of individuals in compartment $j, j=1,2$, at time $t$. By construction $p_1 (t) + p_2 (t) = 1, \forall t$. Then, the evolution of the proportion of infected individuals satisfies the differential equation~:

\begin{eqnarray}
\Frac{dp_2 (t)}{dt} & = & a p_2 (t) p_1 (t) - cp_2 (t) \nonumber \\
&=& ap_2 (t) [1-\Frac{c}{a} - p_2 (t)],
\end{eqnarray}

\noindent where $a>0, c>0.$\vspace{1em}

Parameters $a,c$ are the infinitesimal rates of infection and recovery, respectively. They depend on the time unit, whereas the ratio $a/c$ is independent of the time unit. This explains the role of this ratio in the continuous time epidemiological literature. The change in proportion $p_2 (t)$ is due to i) the rate of new infections $ap_1 (t)$, that involves the proportion of susceptible with a transmission (or contact rate) parameter $a,a>0.$ ii) the recoveries with recovery rate $c>0.$\vspace{1em}

\noindent Let us consider equation (4.1) when~:

\begin{equation}
  \alpha = 1-c/a \in (0,1).
\end{equation}

\noindent If the initial value $p_2 (0)$ is smaller (resp. larger) than $\alpha$, then $p_2 (t)$  starts to increase (resp. decrease), as indicated by the sign of the derivative, up to value $\alpha$. Thus the condition (4.2) is a stability condition for the differential condition (4.1)  with an equilibrium at \footnote{If $\Frac{c}{a} \geq 1, p_2 (t)$ decreases to $0$. At the limit, the disease disappears.} $p_2 (\infty) = \alpha$.\vspace{1em}

\noindent Equation (4.1) can be solved analytically. For instance, if $p_2 (0) < \alpha,$ we get~:

$$
\log \Frac{p_2 (t)}{\alpha - p_2 (t)} - \log \Frac{p_2 (0)}{\alpha - p_2 (0)} = a \alpha t,
$$

\noindent or equivalently,

\begin{equation}
  p_2 (t) = \Frac{\alpha}{1+\Frac{\alpha - p_2 (0)}{p_2 (0)} \exp (- a\alpha t)}.
\end{equation}

\noindent This is a logistic pattern, which is increasing asymptotically up to $\alpha$. Formula (4.3) provides an alternative parametrization of the proportion of infected by means of $(p_2(0), \alpha)$ instead of $(a,c)$. \vspace{1em}

\textbf{Remark 4~:} The SIS model is quite flexible and various extensions of this model have been considered in the literature by introducing a delayed recovery [Cooke, York (1973), Greenberg, Hoppensteadt (1975)], temporary immunity [Xu, Li (2018)], nonlinear incidence rate [Das et al. (2011), Rifhat et al. (2017)], or varying population size [Hethcote, Van den Driessche (1995)].

\subsubsection{Stochastic SIS Model}

The continuous time deterministic model is of limited use for statistical inference for the following reasons. First, the population of interest has a finite size $n$. Second, the available observations are often recorded daily, which is a discrete time setting. Moreover, as the model is deterministic, the statistical theory is not applicable.

A discrete time stochastic analogue of the continuous time deterministic SIS model does not suffer from these limitations. It is based on the analysis of individual histories through Markov chains [see e.g. Gourieroux, Jasiak (2020)]. Let us introduce the following variables:\vspace{1em}

$\bullet N_j (t), j=1, 2$, the counts of individuals in state $j$ at time $t$ with~: $N_1 (t) + N_2 (t) =n.$

$\bullet N_{j|k} (t), j,k=1,2,$ the counts of individuals moving from $k$ to $j$ between $t-1$ and $t$.\vspace{1em}

\noindent  The counts of migrations between the states of infected and susceptible are given below~:

$$
\boxed{
\begin{array}{ll}
  N_{1|1} (t) & N_{2|1} (t) \\ \\
   N_{1|2} (t) & N_{2|2} (t)
\end{array}} \begin{array}{c} N_1 (t-1) \\  \\ N_2 (t-1) \end{array}
$$

\hspace{3,5cm} $N_1 (t)$ \hspace{0,5cm} $N_2 (t)$ \hspace{1cm} $n$\vspace{1em}

\noindent In particular, we get the "conservation of mass idendities" [Matis, Kiffe (2000), Breto et al. (2009)]~:

\begin{equation}
  N_1 (t-1) = N_{1|1} (t) + N_{2|1} (t), N_2 (t-1)
 = N_{1|2} (t) + N_{2|2} (t),
\end{equation}

\noindent and

\begin{equation}
  N_1 (t) = N_{1|1} (t) + N_{1|2} (t), N_2 (t) = N_{2|1} (t) + N_{2|2} (t).
\end{equation}

\noindent We get the following property~:\vspace{1em}

\textbf{Proposition 5~:} Under the Markov chain assumption, conditional on the lagged counts $\left[ N_1 \underline{(t-1)}, \underline{N_2 (t-1)} \right],$ the variables $N_{2|1} (t)$ and $N_{2|2} (t)$ are independent with binomial distributions~:

$$
\mathcal{B} [N_1 (t-1), a N_2 (t-1)/n], \mathcal{B} [N_2 (t-1), 1-c],
$$

\medskip

\noindent where the contagion and recovery parameters take values between 0 and 1.

\medskip

Even though we are still using the notation $a$ and $c$ for the contagion and recovery parameters, these parameters are not identical in the continuous and discrete time. In the discrete time, they are interpreted as daily rates instead of instantaneous rates. This explains the different domains of these parameters, which are $(0, \infty)$ in continuous time 
and $(0,1)$ in discrete time, respectively.

\noindent  It follows that~:\vspace{1em}

\textbf{Corollary 3:} The counts of infected individuals $[N_2 (t)]$ is a Markov process with conditional distribution $\mathcal{B} [n-N_2 (t-1), a N_2 (t-1)/n] \ast \mathcal{B} [N_2 (t-1), 1-c]$, where $\ast$ denotes the convolution of distributions.\vspace{1em}

\noindent The first stochastic feature is due to the finite size of the population. If $n$ tends to infinity, we get approximately:

 $\hat{p}_2 (t) = \Frac{N_2 (t)}{n} \simeq \left( 1-\Frac{N_2 (t-1)}{n}\right) a \Frac{N_2 (t-1)}{n} + \Frac{N_2 (t-1)}{n} (1-c)$.

$$
\begin{array}{l}
  \Longleftrightarrow  \hat{p}_2 (t) \simeq a \hat{p}_2 (t-1) [1-\hat{p}_2 (t-1)] + \hat{p}_2 (t-1) (1-c) \\ \\
  \Longleftrightarrow  \hat{p}_2 (t) - \hat{p}_2 (t-1) = a \hat{p}_2 (t-1) [1- \hat{p}_2 (t-1)]- c \hat{p}_2 (t-1),
\end{array}
$$

\noindent  which is a discrete time analogue of equation (4.1).\vspace{1em}

Let us now discuss the stationarity of process $N_2(t)$. We observe that if $N_2(t-1)=0$, then $N_2(t)=0$. Therefore, the state 0 is an absorbing state of the chain and its stationary distribution is the point mass at 0. However, we are mainly interested in episodes when the disease exists, i.e. $N_2(t)>0$ (and is often large).
For such episodes, we can approximate the distribution in Corollary 3 by $\mathcal{B}^+ [n-N_2(t-1), a N_2(t-1)] \times 
\mathcal{B}^+ [n-N_2(t-1), 1-c]$, where $\mathcal{B}^+$ denotes the binomial distribution restricted to strictly positive values.
Below, $N_2^+(t)$ denotes a Markov chain with such transition distribution.

\medskip

\textbf{Proposition 6:} A sufficient stationarity condition for nondegenerate process $[N_2^+ (t)]$ is $ 0<\alpha = 1-c/a < 1$.\vspace{1em}

\textbf{Proof~:} See Appendix 2.\vspace{1em}

\noindent Considering the process $[N_2(t)]$, the condition given above can be easily explained as follows.

\noindent  Let us assume that the process $\hat{p}_2 (t) = N_2 (t)/n$ is stationary. Since it is bounded, the moments exist and we have~:

\begin{equation}
  E \hat{p}_2 (t) = a E \{ \hat{p}_2 (t) [1-\hat{p}_2 (t)]\} + (1-c) E \hat{p}_2 (t),
\end{equation}

\noindent where $E (\hat{p}_2 (t) [1-\hat{p}_2 (t)]) = \Frac{c}{a} E [\hat{p}_2 (t)]$. This equality implies : $\Frac{c}{a} E \hat{p}_2 (t) \leq E \hat{p}_2 (t).$ Then, the following two cases arise:\vspace{1em}

i) $E \hat{p}_2 (t) = 0$, or equivalently $\hat{p}_2 (t) = 0$, a.s, since $\hat{p}_2 (t)$ is nonnegative.

ii) $\Frac{c}{a} < 1$, which is the condition of Proposition 6.\vspace{1em}

Moreover, we have: $E_{t-1} \hat{p}_2 (t) < (1+a-c) \hat{p}_2 (t-1).$ In our framework $a$ and $c$ are small. If $\Frac{c}{a} > 1, E_{t-1} \hat{p}_2 (t) < \hat{p}_2 (t-1).$ The process $(\hat{p}_2 (t))$ is a nonnegative supermartingale that converges by the Doob's convergence  theorem. Its limit is $\hat{p}_2 (\infty) = 0$ by (4.6). This shows that state $0$ is an absorbing state of the chain, since the conditional distribution of $\hat{p}_2 (t)$ given $\hat{p}_2 (t-1) = 0$ is $\mathcal{P} (0),$ that is a point mass at zero. If $c/a<1$, the contagion is sufficiently large, for any starting value close, but not equal to $0$. By considering $N_2^+(t)$, we ensure that such an escape occurs.
\vspace{1em}

Additional stochastic features can be introduced by considering parameters $\theta_t = (a_t, c_t),$ or $\theta_t = (a_t, \alpha_t)$ varying stochastically in time  [Gray et al. (2011), Dureau et al. (2013), Gourieroux, Lu (2020)]. Then, the model needs to be completed by introducing the conditional distribution of $\theta_t$ given $(\underline{\theta_{t-1}}, \underline{y_{t-1}}) = (\underline{\theta_{t-1}}, \underline{N_2 (t-1)})$. To apply the analysis presented in Section 3, this conditional distribution has to be such that the joint process $(N_2(t), \theta_t)$ is stationary. The dynamic of $\theta_t$ can be either exogenous, or endogenous [see e.g. Ho et al. (2021)]. The latter assumption is suitable to account for changing health policy measures aimed at limiting the transmission (i.e. $a_t$), or controlling the number of hospital beds needed (i.e. $\alpha_t$) [see Wallinga, Teunis (2004)] \footnote{Other stochastic extensions of the SIS model are obtained by replacing the continuous time deterministic models by the associated stochastic differential equations [see e.g. Das et al. (2011), Rhifat et al. (2017), Xu, Li (2018)]. These extensions have less structural epidemiological interpretations. They do not necessarily yield ratios $p_2(t)$ between 0 and 1.}.

\subsection{The Poisson (and Poisson-Gaussian) Approximated SIS Model}

\subsubsection{Poisson approximation}

In practice, the daily infection rate and the recovery rate are small whereas the size of  compartments of infected and suspectible are large. This allows us to approximate the binomial distributions by the Poisson distributions.\vspace{1em}

\textbf{Proposition 7:} Under the regularity conditions for Poisson approximation of the binomial distribution, the conditional distribution of $N_2 (t)$ given $N_2 (t-1)$ is approximately \footnote{The Poisson distribution on the right-hand side of the equality can be replaced by $\mathcal{P}^+$, i.e. the Poisson distribution restricted to strictly positive values, as in Section 4.1. }:

$$
\begin{array}{lcl}
 && \mathcal{P} [[n-N_2 (t-1)] a \Frac{N_2 (t-1)}{n}] * \mathcal{P} [N_2 (t-1) (1-c)] \\ \\
&=& \mathcal{P} [a[n-N_2 (t-1)]  \Frac{N_2 (t-1)}{n}] + (1-c) N_2 (t-1)].
\end{array}
$$

\noindent The advantage of this approximation is twofold : first the convolution produces a closed-form outcome; second we get a dynamic Poisson regression model with lagged endogenous explanatory variables $[n-N_2 (t-1)] \Frac{N_2 (t-1)}{n}$, $N_2 (t-1)$, respectively, and parameters $a, 1-c$. Such a model is a special case of the generalized linear model (GLM or GLIM) that simplifies the analysis of ML estimators [Fahrmeir, Kauffmann (1985), McCullagh, Nelder (1989) for introduction to GLIM, Cai, Fan, Li (2000) for varying coefficients in GLIM].

\noindent To clarify the link between the notation of the SIS model and the notation in Sections 2 and 3, let us define: $y_t = N_2 (t), \theta = (a,1-c)',$

$$
z_{t-1} = [(n-y_{t-1}) y_{t-1}/n, y_{t-1}]'\; \mbox{and}\; \lambda_{t-1} (\theta) = z'_{t-1} \theta.
$$

\noindent Then the pseudo-log-likelihood function for the time series of counts is:

$$
\begin{array}{lcl}
  \log l(y_t | y_{t-1} ; \theta) &\propto & -\lambda_{t-1} (\theta) + y_t \log \lambda_{t-1} (\theta) \\ \\
  &=&-z'_{t-1} \theta + y_t \log (z'_{t-1} \theta).
\end{array}
$$

\noindent Its first-and second-order derivatives  are~:

$$
\begin{array}{lcl}
L_{1t} (\theta) & = & z_{t-1} \left[ \Frac{y_t}{z'_{t-1} \theta} - 1\right] = z_{t-1}
                                     \Frac{v_t (\theta)}{z'_{t-1} \theta},\; \mbox{with}\; v_t (\theta) = y_t - z'_{t-1} \theta, \\ \\
     L_{2t} (\theta) & = & \Frac{-z_{t-1} z'_{t-1} y_t}{(z'_{t-1} \theta)^2}.
\end{array}
$$

\noindent The first-order conditions for the TLML estimator:

$$
\Sum^{T-1}_{h=0} \{ w(h) z_{T-h-1} \left[ \Frac{y_{T-h}}{z'_{T-h-1} \hat{\theta}_T (w)} - 1\right] = \Sum^{T-1}_{h=0} \left\{ w(h) \Frac{z_{T-h-1}}{z'_{T-h-1} \hat{\theta}_T (w)} v_{T-h} [\hat{\theta}_T (w)]\right\} = 0,
$$

\noindent are weighted  conditions of orthogonality between the explanatory variables and the residuals $v_t [\hat{\theta}_T (w)].$

These first-order conditions are nonlinear in $\hat{a}_T (w), \hat{c}_T (w)$ and do not produce closed form expressions of the TLML estimators. In particular, their nonlinearity likely induces biases so that $\hat{a}_T (w) - a_T$ is not zero mean.\vspace{1em}

\textbf{Remark 5~:} Another difficulty is due to the form of lagged endogenous regressors for $a$ and $c$. Indeed, if $N_2 (t-1)/n$ is small the regressors are $N_2 (t-1)$ for parameter $c$, $N_2 (t-1) [1-\Frac{N_2 (t-1)}{n}]$ for $a$, and this second regressor is close to $N_2 (t-1)$. Therefore there is a problem of quasi-collinearity that can render accurate joint estimation of parameters difficult.
\vspace{1em}

In the framework of Poisson approximation, the local maximum likelihood approach can be compared with the rolling regression approach used in epidemiology [see e.g. Cori et al. (2013), Calafiore et. al. (2020), Rubio, Herrero, Wang (2021) \footnote{ Also used with rolling over 1 day only [see, e.g. Waqas et al. (2020)]. In applications to COVID-19, the rolling methods are used for the SIR (Susceptible-Infected-Recovered) and SIRD (D for Deceased) models.}]. Under this approach, we consider a time discretized version of equation (4.1), that is:

\begin{eqnarray*}
\lefteqn{\frac{N_2(t)}{n} - \frac{N_2(t-1)}{n} \approx a [ 1- \frac{N_2(t-1)}{n}] \frac{N_2(t-1)}{n} - c \frac{N_2(t-1)}{n}} \\ 
& \iff & N_2(t) \approx N_2(t-1) [ 1 -\frac{N_2(t-1)}{n} ] + (1-c) N_2(t-1).
\end{eqnarray*}

\noindent Then, the estimators of parameters $a$ and $c$ are obtained from OLS regressions applied by rolling \footnote{Sometimes these rolling regressions are analysed by using Bayesian methods, as in Cori et al. (2013), instead of the frequentist approach.}.
In comparison with the local maximum likelihood based on the Poisson approximation, this approach based on rolling OLS regressions disregards the conditional distribution of the errors, and in particular, their conditional heteroskedasticity.  

\subsubsection{Poisson-Gaussian approximation}

When the intensity parameter $\lambda$ is large, the Poisson distribution $\mathcal{P} (\lambda)$ can be approximated by the Gaussian distribution $N (\lambda, \lambda)$.\vspace{1em}

\textbf{Proposition 8:} If the intensities $\lambda_{t-1} (\theta)$ are large, the conditional distribution of $y_t = N_2 (t)$ given $y_{t-1} = N_2 (t-1)$ is approximately the Gaussian distribution $N[\lambda_{t-1} (\theta), \lambda_{t-1} (\theta)].$\vspace{1em}

\noindent This Gaussian distribution with the mean equal to the variance is henceforth referred to as Poisson-Gaussian.  The dated Poisson-Gaussian pseudo-log-likelihood is~:

$$
\log l(y_t | y_{t-1} ; \theta) \propto - \Frac{1}{2} \log \lambda_{t-1} (\theta) - \Frac{1}{2} \Frac{y^2_t}{\lambda_{t-1} (\theta)} + \Frac{1}{2} \lambda_{t-1} (\theta)
$$

$$
= -\Frac{1}{2} \log (z'_{t-1} \theta) - \Frac{1}{2} \Frac{y^2_t}{z'_{t-1} \theta} + \Frac{1}{2} z'_{t-1} \theta.
$$

\noindent Its first-order derivatives  are~:

$$
\begin{array}{lcl}
L_{1t} (\theta) & = & \Frac{1}{2 (z'_{t-1} \theta)^2} z_{t-1} [y^2_t - z'_{t-1} \theta + (z'_{t-1} \theta)^2] \\ \\
&=& \Frac{1}{2 (z'_{t-1} \theta)^2} z_{t-1} v_t (\theta),
\end{array}
$$

\noindent where $v_t (\theta) = y^2_t - E_\theta (y^2_t | y_{t-1})$. Therefore the FOC for TLML~:

$$
\Sum^{T-1}_{h=0} \{\Frac{ w(h)}{[z'_{T-h-1} \hat{\theta}_T (w)]^2} z_{T-h-1} v_{T-h} (\hat{\theta}_T (w))\} = 0,
$$

\noindent are still weighted nonlinear conditions of orthogonality between the explanatory variables and the residuals.

\subsection{Numerical Illustration}
\subsubsection{The Design}

We consider the stochastic SIS model of Proposition 1, where the recovery rate is constant and  the contagion (transmission) parameter is such that~:

\begin{equation}
  \log a_t - \log a^* = \rho (\log a_t - \log a^*) + \sigma u_t,
\end{equation}

\noindent where the errors $u_t$ are i.i.d.  N(0,1). The length of the period considered is 600 days.

\noindent The starting values of the count processes and stochastic contagion are fixed and given below:

$$
N_1 (0) = 4915, N_2 (0) = 85, n=N_1 (0) + N_2 (0) = 5000, a_0 = 0.2.
$$

\noindent The long run parameters are fixed to: $a^*=0.2, c=0.98 a=(0.98) 0.2$ and then $\alpha = 1-c/a=0.02$ to satisfy the stationarity condition in Proposition 2.

\noindent The parameters $\rho, \sigma$ of the contagion dynamics are~: \vspace{1em}

(i) $\rho = 0, \sigma = 0,$ which corresponds to constant contagion $a_t=0.2, \forall t$.


ii) $\rho = 0.99, \sigma = 0.01$, which corresponds to an ULR contagion process and a smooth stochastic evolution of $a_t$.\vspace{1em}

\noindent We apply the Poisson TLML with three different one-sided geometric weighting schemes with geometric rate $w$ equal to 0.1, 0.5 and 0.9.

\subsubsection{Constant Contagion Parameters}

Figure 1 displays a joint simulated path of the counts of infected $N_2 (t)$ (solid black line), new infected  $N_{2|1} (t)$ (red dashed line), and new recovered individuals $N_{1|2} (t)$ (dotted green line).\vspace{1em}

[Insert Figure 1 : Trajectory of Counts, $a$ Constant]\vspace{1em}

These trajectories resemble the trajectories of stationary processes. They feature pseudo-cycles and spikes due to the nonlinear dynamic that underlies the SIS model. Such trajectories are compatible with the simulations reported in the epidemiological literature [see e.g. Das et al. (2011), Figures 1.b, 2.b].

Even though migration counts $N_{1|2}, N_{2|1}$  are of similar magnitude, these variables occasionally display  different cluster effects, that may lead to significant variations in the counts of infected, computed by cumulating these migration counts.

We examine the behavior of  TLML estimates based on the simulated data in Figure 1. The parameters of interest are the daily contagion rate $a$ (with the true value $a=0.2$) and the daily "reproductive number" $R_0 = a+1-c$ (with the true value $R_0=1$). When the epidemics is such that $N_2(t)/n$ is small, as in COVID-19 for example, the conditional distribution of $N_2(t)$ is approximately Poisson $\mathcal{P}[(1+a-c)N_2(t-1)] = \mathcal{P} [R_0 N_2(t-1)]$. 
$R_0$ can be interpreted  as a rate of explosion or a reproductive number. There exist alternative definitions of the reproductive number in the literature, such as $R^* = a/c$ for the deterministic models in continuous time. Our definition has been adjusted for the stochastic and discrete time features. If $R_0 >1$, there is explosive growth in the mean, and a decline, otherwise.

Figure 2 below displays the dynamics of the estimated contagion parameter for the three weighting schemes considered. The dynamics of the estimated reproductive number  for the three weighting schemes considered 
is plotted in Figure 3.\vspace{1em}

\medskip

[Insert Figure 2: Estimates of Contagion Parameter, $a$ Constant]\vspace{1em}

\medskip

[Insert Figure 3: Estimates of the Reproductive Number $R_0$, $a$ Constant]\vspace{1em}

\medskip

\noindent As expected, the trajectories are less erratic for higher values of $w$. However, even for $w=0.9$ one can observe patterns, which are likely due to the nonlinear dynamics of the SIS model and/or the local analysis. More specifically, the TLML estimator of the contagion parameter $a$ is slightly overestimated. Also, at times close to the end of the sampling period when $N_2(t)$ increases sharply, we observe contagion parameter values close, or equal to the upper bound of 1.

 While the path of the estimated contagion parameter $a$ is smooth for $w=0.9$, it is not the case for the reproductive number, which is more impacted by the peaks in the counts of infected displayed in Figure 1. It is easy to check that the estimated reproductive number is close to a local measure of the gross rate of increase of infected, equal to a weighted average of ratios $N_2(t)/N_2(t-1)$.

Figure 4 provides the sample density of $\hat{a}_t - a =  \hat{a}_t  - 0.2$, which is a nonparametric estimate of the stationary distribution of $\hat{a}_t - a$ calculated from the parameter estimates reported above, after removing the extremes.

\medskip

[Insert Figure 4: Deviation $\hat{a}_t  - 0.2$, $a$ Constant] \vspace{1em}

\noindent We observe that the density of the estimator of the contagion parameter is centered at a value slightly greater than 0, which confirms a slight bias observed in Figure 2. The stationary densities become more peaked when $w$ increases. 

The density of the deviation of reproductive number estimator from 1 is centered at 0, as shown in Figure 5 below. 

\medskip

[Insert Figure 5:  Deviation $\hat{R}_{0,t} - 1$, $a$ Constant] \vspace{1em}

\medskip

\noindent The stationary densities become more peaked when  $w$ increases, similarly to the pattern observed in Figure 4.  

\medskip

Table 1 below shows the summary statistics for the series of deviations $\hat{a}_t - 0.2$ and $\hat{R}_{0,t} - 1$ corresponding to the constant contagion parameter $a=0.2$.

\begin{table}[ht]
\caption{Deviations-Summary Statistics: Constant $a$}
\begin{center}
\begin{tabular}{cccc|cccc}
    \hline
   \multicolumn{4}{c}{$\hat{a}_t - 0.2$} & \multicolumn{4}{|c}{$\hat{R}_{0,t} - 1$} \\
   \hline
    mean & s.d. & skew    &  kurt  & mean & s.d. & skew    &  kurt \\
    \hline 
    \multicolumn {8}{c}{w=0.1}\\
0.027 &  0.035 & 0.283 & 2.898 &  0.005 & 0.069 & 0.198 & 2.670 \\
\hline
\multicolumn {8}{c}{w=0.5}\\
0.029 & 0.032 & 0.379 &  3.252 & 0.008  & 0.063  & 0.255 & 2.946 \\
\hline
\multicolumn {8}{c}{w=0.9}\\
0.026 & 0.007 &  0.194 & 2.615 &  0.003 &  0.015 &  0.114 & 2.669 \\ 
    \hline
\end{tabular}
\end{center}
\end{table}

\medskip

The deviations $\hat{a}_t - 0.2$  have means close to 0.02 and are slightly skewed with  kurtosis values close to 3. The deviations $\hat{R}_{0,t} - 1$, have means closer to 0, are more dispersed, less skewed with slightly lower kurtosis values. The lowest dispersion is obtained for $w=0.9$.

Due to the quasi-collinearity issue pointed out earlier and the spikes in the trajectory of $N_2(t)$, the 
estimation of parameters  $a$ and $c$  can be  computationally challenging. These estimators occasionally hit the bounds of 0 and 1, respectively. This explains the presence of estimated values $\hat{a} =1$, which are marked by red dots in Figure 2. To illustrate the quasi-collinearity, Figure 12 in Appendix 3 displays the eigenvalues of the sample information matrix of the weighted local log-likelihood $J_{\infty}$. We observe that at least one of the eigenvalues is close to 0 for most of the times.

\subsubsection{Stochastic Contagion}

Figure 6 displays a joint simulated path of the counts of infected $N_2 (t)$ (solid black line), new infected  $N_{2|1} (t)$ (red dashed line), and new recovered individuals $N_{1|2} (t)$ (dotted green line) for a stochastically time varying contagion parameter.\vspace{1em}

\medskip

[Insert Figure 6 : Trajectory of Counts, Stochastic $a$ ]\vspace{1em}

\noindent By introducing a stochastic contagion, we eliminate the Markov property of the count process
and introduce more persistence. In this  nonlinear dynamic framework this creates larger peaks and throughs, as compared with the trajectory in Figure 1. 

The path of the stochastic contagion parameter $a_t$ is displayed in Figure 7 below.

\medskip

[Insert Figure 7 : Trajectory of Stochastic $a$ ]\vspace{1em}

\noindent The patterns in $a_t$ affect the dynamics of counts of infected. For example, any decline in 
$a_t$ is associated with an expected  decline in $N_2(t)$.
 
The functional estimator of contagion parameter is shown in Figure 8 for the three weighting schemes. As in the case of constant $a$, the estimator has a slight positive bias and lies above the path of the true $a_t$ plotted in red. Nevertheless, the first few values of the estimator lie on the red line for each weighting scheme, indicating that the estimator  $\hat{a}_t$ is unbiased of $a_t$ when the count $N_2(t)$ is close to 0 at time $t$=100.

\medskip

[Insert Figure 8 : Trajectory of Contagion Parameter Estimates, Stochastic $a$ ]\vspace{1em}

\noindent The path of the functional contagion estimator is smoother for higher values of $w$. 

Figure 9 shows the estimated values of the reproductive number $R_0$.

\medskip

[Insert Figure 9 : Estimates of Reproductive Number $R_{0,t}$, Stochastic $a$ ]\vspace{1em}

\noindent  We observe that, unlike the estimator of $a_t$, the estimator $\hat{R}_{0,t}$ has the highest bias when  
$N_2(t)$ is close to 0, which is due to an increased bias in $\hat{c}_t$. 
Overall, the estimate $\hat{R}_{0,t}$ of the reproductive number is close to the true value $R_{0,t}$, except around times $t=100$ and $t=400$ following a decrease in $N_2(t)$.

Figures 10 and 11 show the sample densities of $\hat{a}_t -a_t$ and   $\hat{R}_{0,t} - R_{0,t}$.

\medskip

[Insert Figure 10 : Deviation $\hat{a}_t -a_t$, Stochastic $a$ ]\vspace{1em}

\medskip

[Insert Figure 11 : Deviation $\hat{R}_{0,t} -  R_{0,t}$, Stochastic $a$ ]\vspace{1em}

\noindent Figure 10 confirms the "slight bias" in estimator $\hat{a}_t$, as the densities of deviations from the true $a_t$ are centered at a value slightly above 0. Figure 11 shows that the estimator of the reproductive number is unbiased, although the density of the estimator for $w=0.9$ has a heavy left tail.

Table 2 gives the summary statistics for the series of deviations $\hat{a}_t - a_t$ and $\hat{R}_{0,t} - R_{0,t} $ corresponding to the case of time varying stochastic contagion parameter $a_t$.

\begin{table}
\caption{Deviations-Summary Statistics: Stochastic $a$}
\begin{center}
\begin{tabular}{cccc|cccc}
    \hline
   \multicolumn{4}{c}{$\hat{a}_t - a_t$} & \multicolumn{4}{|c}{$\hat{R}_{0,t} - R_{0,t}$} \\
   \hline
     mean & s.d. & skew    &  kurt  & mean & s.d. & skew    &  kurt \\
    \hline 
    \multicolumn {8}{c}{w=0.1}\\
0.028 &  0.033 &  0.061 &  5.084 & 0.007 &  0.062 &  0.121 & 3.128  \\
\hline
\multicolumn {8}{c}{w=0.5}\\
0.027 &  0.022  & 0.487 &  6.117 &  0.007 &  0.042 &  0.058 & 4.264\\
\hline
\multicolumn {8}{c}{w=0.9}\\
0.026 &   0.010 &  2.812 & 34.946 &  0.005 &  0.017 &  -0.163 & 3.351   \\ 
    \hline
\end{tabular}
\end{center}
\end{table}

The deviations $\hat{a}_t - a_t$ have means close to 0.2, and positive skewness measures, except for 
$w=0.1$. Their densities are heavy tailed for all weights. The deviations $\hat{R}_{0,t} - R_{0,t} $ have means close to 0, are slightly skewed, less dispersed and have tails close to the normal. By comparing with Table 1, we find that the main effect of stochastic contagion is the increase of tails of estimation errors.

The computational difficulty resulting in some estimator values $\hat{a}_t$ being close to 1 is illustrated in Appendix 3, Figure 13, which displays the eigenvalues of the sample information matrix. Similarly to the case of constant $a$, one eigenvalue takes a value close to 0 at almost all times t, due to quasi-collinearity.

It would be possible to improve the approximation of the contagion parameter and reproductive number, if additional information on the recovery rate $c$ is available. For instance, if infected individuals are hospitalized, the average time spent in the hospital can provide an approximation of $1/c$. Then, at date $t$ the estimation of $a_t$ can be performed given a proxy of parameter $c_t$  equal to the inverse of the average hospitalization time. Such optimization of the weighted log-likelihood would be carried out with respect to parameter $a$ only, although the first-order conditions are still nonlinear in $a$. This approach can greatly reduce or circumvent the quasi-collinearity issue and even eliminate the bias of $\hat{a}_t$.

\section{Concluding Remarks}

The weighted estimation methods are often applied by rolling to predict the parameters of interest over a period of time. This approach can be used for the analysis of a disease transmission in a population. Then, the  model is a nonlinear dynamic model, which can accommodate, even under stationarity, occasionally occurring peaks, clustering, tipping points, chaotic effects and other nonlinear dynamic features.

We have discussed the analytical properties of the TLML in order to explain the potential biases induced by this approach, which can be due to the nonlinear first-order conditions, local analysis, or presence of dynamic heterogeneity.

The results have been illustrated by a simulation study of the SIS epidemiological model. In this simple framework, we have revealed the lack of robustness and accuracy of the approach [also observed in other epidemiological models and estimation approaches, see e.g. Elliot, Gourieroux (2021)]. More accurate results could be obtained by estimating the contagion given a proxy of the recovery rate based on an auxiliary estimator.

\newpage
  \begin{center}
\textbf{Appendix 1}\vspace{1em}

\textbf{Second-Order Expansion}\vspace{1em}
\end{center}
     \setcounter{equation}{0}\def\theequation{a.\arabic{equation}}

    Even though the TLML estimator is consistent, we expect the  bias adjustment formula to be non-standard. We consider below the second-order expansion of the FOC [see Gosh, Subramanyan (1974), Efron (1975), Firth (1993), Kosmidis, Firth (2009)]. For expository purpose, we assume $\dim \theta_t = 1$ and give the general formulas at the end of this Appendix. We define:

    $$
    \begin{array}{lcl}
    L_{1t} (\theta^*_\infty) & = & \Frac{\partial \log l}{\partial \theta} [y_t |y_{t-1}; \theta^*_\infty], \\ \\
    L_{2t} (\theta^*_\infty) & = & \Frac{\partial^2 \log l}{\partial \theta^2} [y_t |y_{t-1}; \theta^*_\infty], \\ \\
    L_{3t} (\theta^*_\infty) & = & \Frac{\partial^3 \log l}{\partial \theta^3} [y_t |y_{t-1}; \theta^*_\infty].
    \end{array}
    $$

\noindent The second-order expansion of the FOC in a neighbourhood of the pseudo-true value is~:

$$
\Sum^{T-1}_{h=0} [w(h) L_{1,T-h} (\theta^*_\infty)] + \Sum^{T-1}_{k=0} [w(h) L_{2,T-h} (\theta^*_\infty)] (\hat{\theta}_T (w) - \theta^*_\infty)
$$

 $$
 + \Frac{1}{2} \Sum^{T-1}_{k=0} [w(h) L_{3,T-h} (\theta^*_\infty)] (\hat{\theta}_T (w) - \theta^*_\infty)^2 = o_P,
 $$

\noindent or equivalently,

\begin{eqnarray}
  &&\Sum^{T-1}_{h=0} [w(h) L_{1,T-h} (\theta^*_\infty)] + W_T E_0 L_{2,t} (\theta^*_\infty) (\hat{\theta}_T (w) - \theta^*_\infty)\nonumber \\
  &+& \Sum^{T-1}_{h=0} \{ w(h) [L_{2,T-h} (\theta^*_\infty) - E_0 L_{2,t} (\theta^*_\infty)]\} (\hat{\theta}_T (w) - \hat{\theta}^*_\infty)\nonumber \\
  &+& \Frac{1}{2} W_T E_0 L_{3,t} (\theta^*_\infty) (\hat{\theta}_T (w) - \theta^*_\infty)^2 ) = o_P.
\end{eqnarray}

\noindent  From (3.9), it follows that:

$$
\hat{\theta}_T (w) - \theta^*_\infty = \Frac{I_T (w;\theta^*_\infty)^{1/2}}{W_T J(\theta^*_\infty)} X_T + o_P,
$$

\noindent where $X_T \simeq N (0;Id).$\vspace{1em}

\noindent This expression can be plugged into the two last terms of the second-order expansion to get~:

$$
\begin{array}{lcl}
  &&\Sum^{T-1}_{h=0} [w(h) L_{1,T-h} (\theta^*_\infty)] = W_T J(\theta^*_\infty) (\hat{\theta}_T (w) - \theta^*_\infty) \\ \\
  &+& \Sum^{T-1}_{k=0} \{ w(h) [L_{2,T-h} (\theta^*_\infty) + J (\theta^*_\infty)] \Frac{I_T (w;\theta^*_\infty)^{1/2}}{W_T J (\theta^*_\infty)} X_T \\ \\
  &+& \Frac{1}{2} W_T E_0 L_{3,t} (\theta^*_\infty) \Frac{I_T (w;\theta^*_\infty)}{W^2_T J(\theta^*_\infty)^2} X^2_T = o_P.
\end{array}
$$

\noindent Let us consider the variable~:

$$
Z_T = \left[ \Sum^{T-1}_{h=0} \Sum^{T-1}_{k=0} w(h) w(k) I_2 (h-k;\theta^*_\infty)\right]^{-1/2} \Sum^{T-1}_{h=0}
[w(h) \Frac{\partial^2 \log l}{\partial \theta^2} (y_{T-h} | y_{T-h-1} ; \theta^*_\infty)]
$$

$$
\equiv I_{2,T} (w;\theta^*_\infty)^{-1/2} \Sum^{T-1}_{h=0} (w(h) \Frac{\partial^2 \log l}{\partial \theta^2} (y_{T-h}|y_{T-h-1}; \theta^*_\infty)],
$$

\noindent where $I_2 (h;\theta^*_\infty) = Cov_0 \left[ \Frac{\partial^2 \log l}{\partial \theta^2} (y_t | y_{t-1};\theta^*_\infty), \Frac{\partial \log l}{\partial \theta^2} (y_{t-h} |y_{t-h-1};\theta^*_\infty)\right].$\vspace{1em}

\noindent Then we get:

$$
\begin{array}{lcl}
 && \hat{\theta}_T (w) - \theta^*_\infty  =  \Frac{I_T (w;\theta^*_\infty)^{1/2}}{W_T J (\theta^*_\infty)} X_T \\ \\
 & - & \Frac{I_T (w;\theta^*_\infty)^{1/2}}{W_T J (\theta^*_\infty)} \Frac{I_{2T} (w;\theta^*_\infty)^{1/2}}{W_T  J (\theta^*_\infty)} X_T Z_T \\ \\
 &-& \Frac{1}{2} \Frac{E_0 L_{3,t} (\theta^*_\infty)}{J(\theta^*_\infty)} \Frac{I_T (w;\theta^*_\infty)}{W^2_T J (\theta^*_\infty)^2} X^2_T + o_P.
\end{array}
$$

This expansion provides an approximation of the difference between the TLML estimator and the pseudo-true value as a quadratic function of the pair $(X_T, Z_T),$ which is asymptotically normally distributed with zero mean components with variances $V X_T \simeq V Z_T \simeq Id$ and a non-zero correlation, in general. Alternative bias adjustments could be based on the pseudo score [Lambrecht et al. (1997)].

\newpage
  \begin{center}
\textbf{Appendix 2}\vspace{1em}

\textbf{ Stationarity Condition}\vspace{1em}
\end{center}
     \setcounter{equation}{0}\def\theequation{a.\arabic{equation}}

     \textbf{a) Condition for ergodicity}\vspace{1em}

\noindent Whenever the Markov chain is irreducible, we can use the sufficient conditions for ergodicity provided in Tweedie (1971). They are~:\vspace{1em}

     i) $E [\hat{p}_2 (t) | \hat{p}_2 (t-1)]$ is bounded.

     ii) $E [\hat{p}_2 (t) - \hat{p}_2 (t-1) | \hat{p}_2 (t-1) = x] \leq - \varepsilon$, for $x$ outside a compact set.\vspace{1em}

\noindent Condition i) is satisfied since $\hat{p}_2 (t) < 1$.\vspace{1em}

\noindent  Let us now consider condition ii). We have~:

     $$
     \begin{array}{l}
     E [\hat{p}_2 (t) - \hat{p}_2 (t-1) | \hat{p}_2 (t-1) = x] \\ \\
     = a x(1-x) - c x <-cx.
     \end{array}
     $$

\noindent Therefore condition ii) is satisfied with the compact set $K = [\varepsilon/c,1]$, and $\varepsilon < c$.\vspace{1em}

     \textbf{b) Condition for irreducibility}\vspace{1em}

\noindent The different types of behaviour discussed following Proposition 6 are related to the irreducibility properties of the Markov chain [see e.g. Rio (2017), Chapter 9, and Chotard, Auger (2019)]. This irreducibility is due to the condition $0< 1-c/a <1$, and the replacing of $\mathcal{B}$ by $\mathcal{B}^+$ (or $\mathcal{P}$
by $\mathcal{P}^+$) in the definition of the chain.

\newpage

\begin{center}
\textbf{Appendix 3}\vspace{1em}

\textbf{Additional Figures}\vspace{1em}
\end{center}
\vspace{1em}

\textbf{3.1. Quasi-Collinearity}\vspace{1em}

To provide some insights on quasi-collinearity, we compute the sample information matrix from the Hessian of the temporally local log-likelihood function and report its eigenvalues. As expected these eigenvalues are positive and one of these  eigenvalues is small and close to 0.

\medskip

[Insert Figure 12: Eigenvalues of Estimated Information Matrix: Constant $a$ ]

\medskip

[Insert Figure 13: Eigenvalues of Estimated Information Matrix: Stochastic $a$ ]

\bigskip

\textbf{3.2 Persistence of Estimates and Stochastic Parameters}\vspace{1em}

Additional summary statistics provided below concern the joint dynamics of the stochastic parameters of interest and their local estimates. They are provided in Figures 14 and 15 for the contagion parameter and reproductive number, respectively. For each value of $w$ considered, the diagonal panels show the autocorrelations of the estimates and of the stochastic parameters, respectively. The off-diagonal panels show the cross-correlations between the estimates and stochastic parameters. 

The autocorrelation function (ACF) of $a_t$ and $R_{0,t}$ reveals the local-to-unity property of the ULR 
process. For $w=0.1$, the weighted estimator is more global and appears almost uncorrelated with the underlying stochastic parameter. We observe that the cross-correlations, which often take small values, do not decay to zero with the lag. This is a consequence of the ULR $a_t$ process. For larger $w$,  the estimates display more persistence, although the autocorrelations of $\hat{a}_t$ are smaller and decay faster than those of $a_t$.

\medskip

[Insert Figure 14: Joint ACF of $a_t$ and $\hat{a}_t$]

\medskip

[Insert Figure 15: Joint ACF of $R_{0,t}$ and $\hat{R}_{0,t}$]

\bigskip

\textbf{3.3 Nonlinear Prediction Performance}\vspace{1em}

Under the standard maximum likelihood approach, the values of the log-likelihood at the optimum can be used to perform tests based on the likelihood ratios, especially of the time varying hypotheses $H_{0T} = \{ R_{0T} >1 \}$. It can be used to check the "joint" accuracy of $\hat{a}_t, \hat{c}_t$, as a global prediction measure. In this respect, Figure 16 displays the evolution of the local weighted  log-likelihood. This evolution has to be compared with the trajectory of counts in Figure 6. As expected, the local nonlinear fit is better during the episodes of counts evolving without sudden jumps, either corresponding to "local" trends, or rather stable evolution. The values of the log-likelihood are decreasing in the neighbourhoods of extreme peaks or throughs.

Alternative measures of prediction performance could also be constructed by comparing at each date the observed number of infected individuals $N_2(t)$ with its estimator-based prediction: 

$$\hat{N}_2(t) = \hat{a}_t [n- N_2(t-1)]
\frac{N_2(t-1)}{n} + (1-\hat{c}_t)N_2(t-1).$$

\noindent The difference $N_2(t) - \hat{N}_2(t)$ depends on the theoretical prediction error (which is present when the 	true parameter values are used) and the estimation error due to replacing the true parameters by their estimates in the theoretical pointwise prediction.

\begin{center}
\begin{figure}
\centering
\includegraphics[width=14cm,height =14cm,angle = 0]{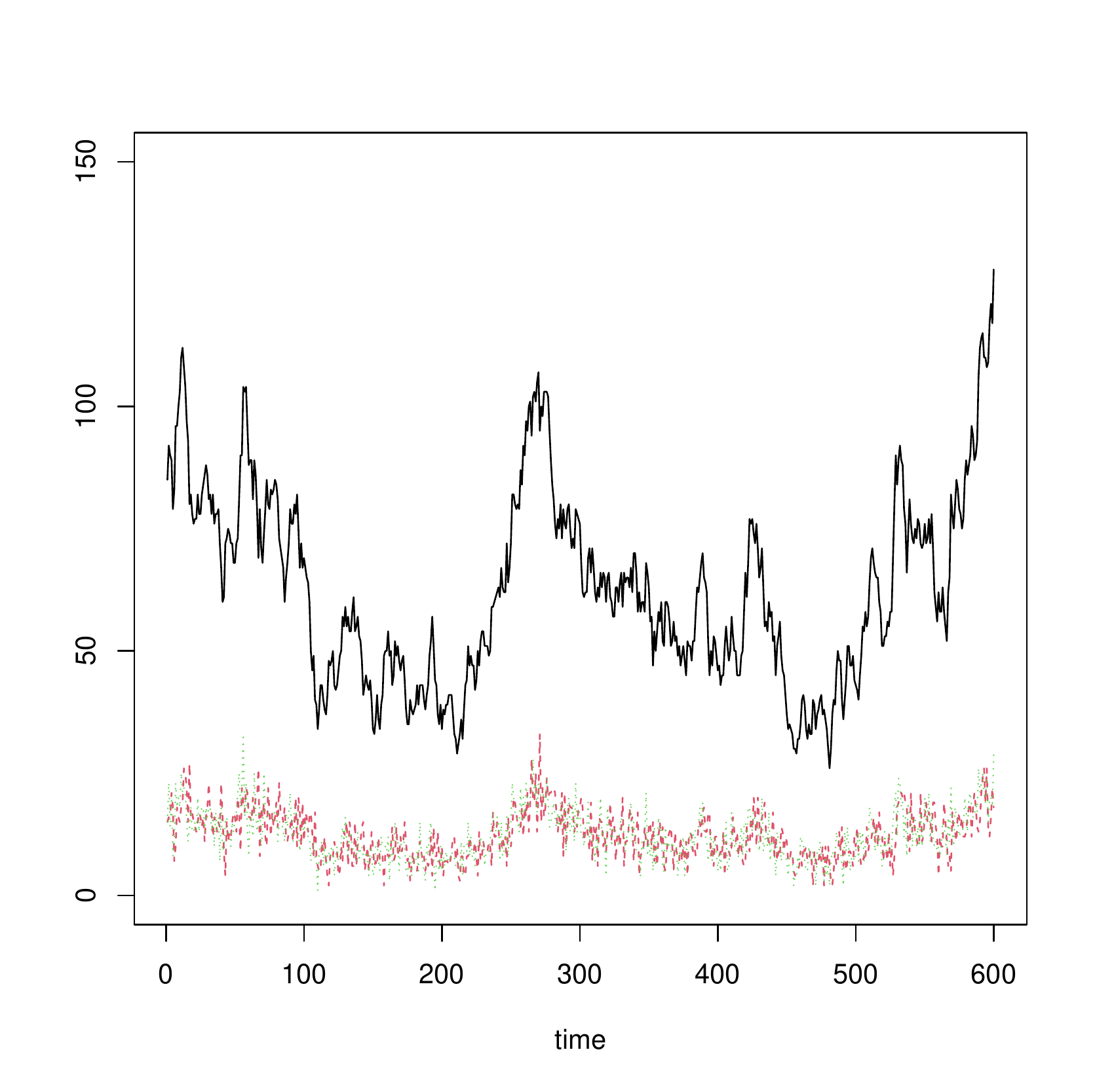}
{\caption*{Figure 1: Trajectories of Counts, $a$ Constant   }}
\end{figure}
\end{center}

\begin{figure}
\centering
  \begin{subfigure}[t]{\textwidth}
  \centering
  \includegraphics[height=2.2in, width=4.4in]{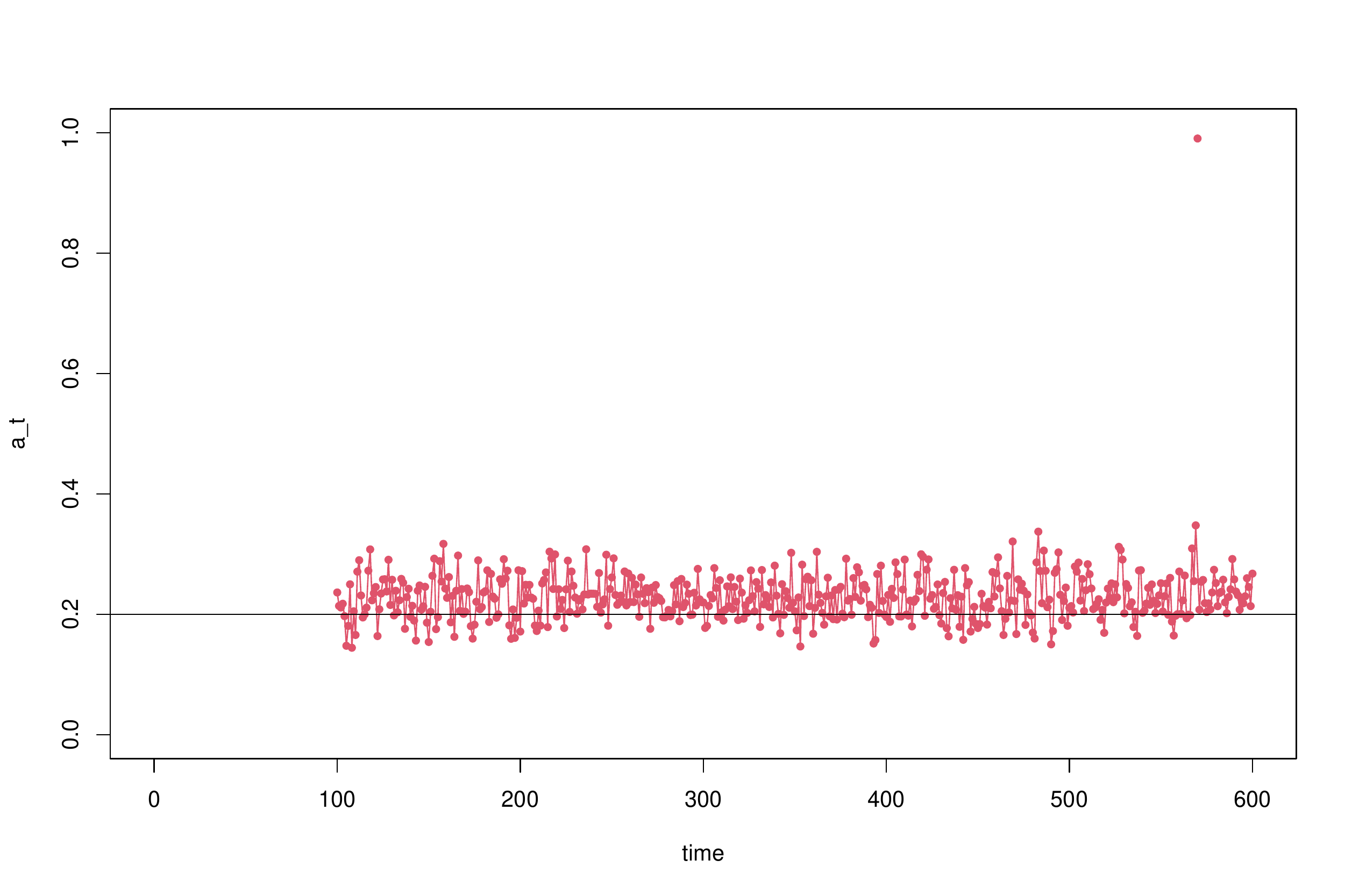}
    {\caption*{w=0.1}}
  \end{subfigure}
    \begin{subfigure}[t]{\textwidth}
  \centering
 \includegraphics[height=2.2in, width=4.4in]{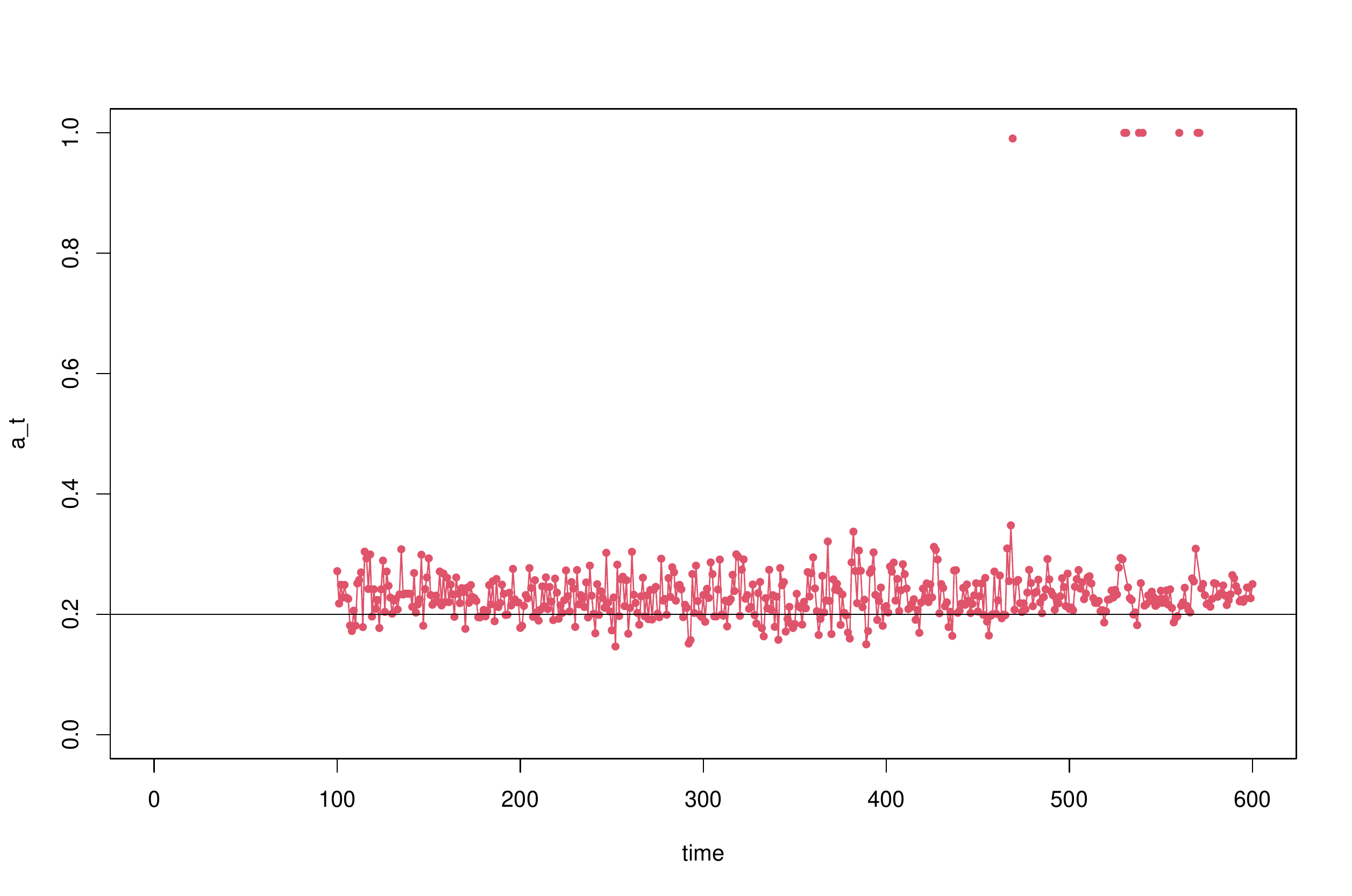}
  {\caption*{w=0.5}}
  \end{subfigure}
   \begin{subfigure}[t]{\textwidth}
  \centering
 \includegraphics[height=2.2in, width=4.4in]{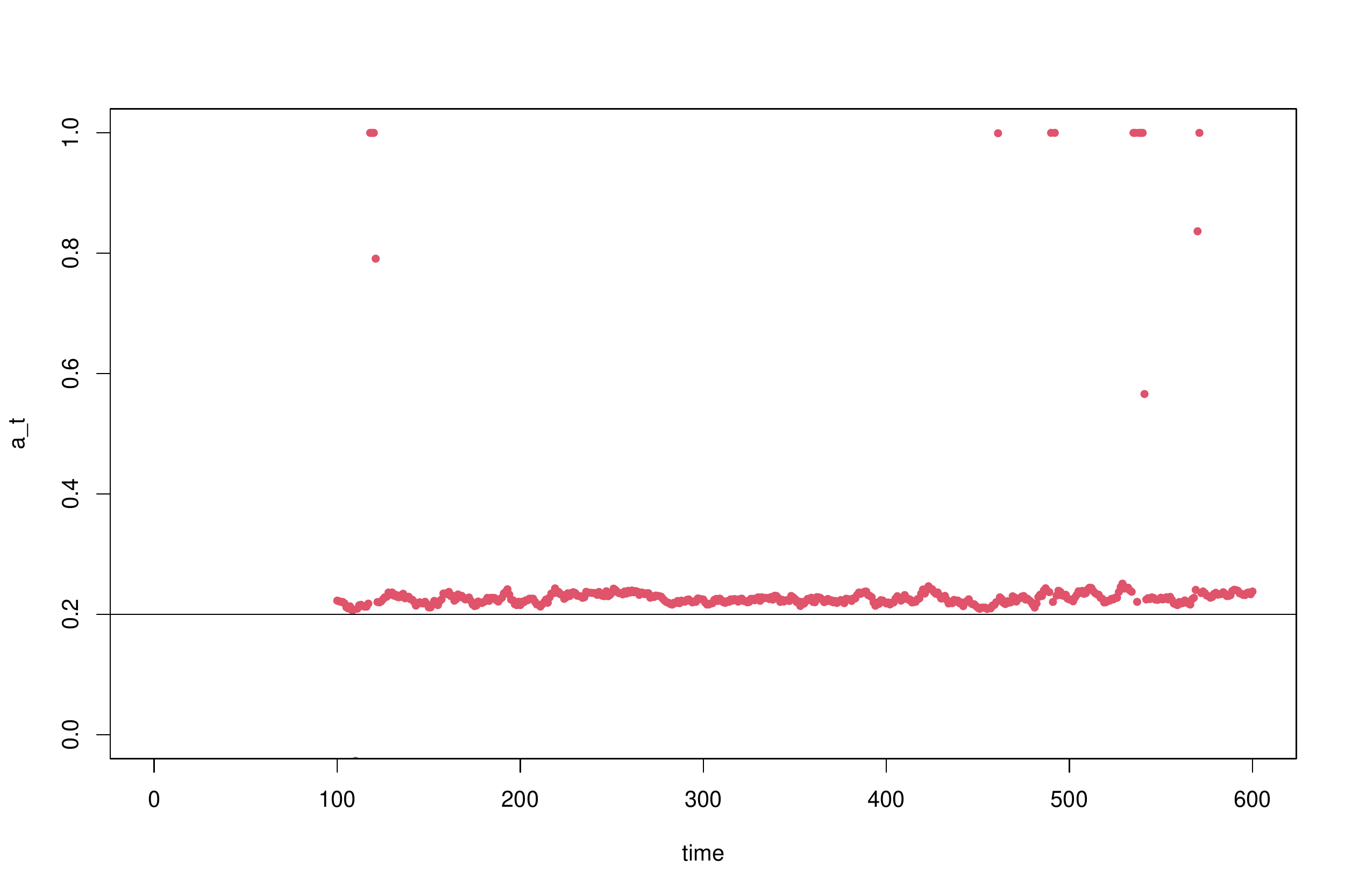}
 {\caption*{w=0.9}}
  \end{subfigure}
 Figure 2: Estimates of Contagion Parameter, $a$  Constant
\end{figure}

\begin{figure}
\centering
  \begin{subfigure}[t]{\textwidth}
  \centering
\includegraphics[height=2.2in, width=4.4in]{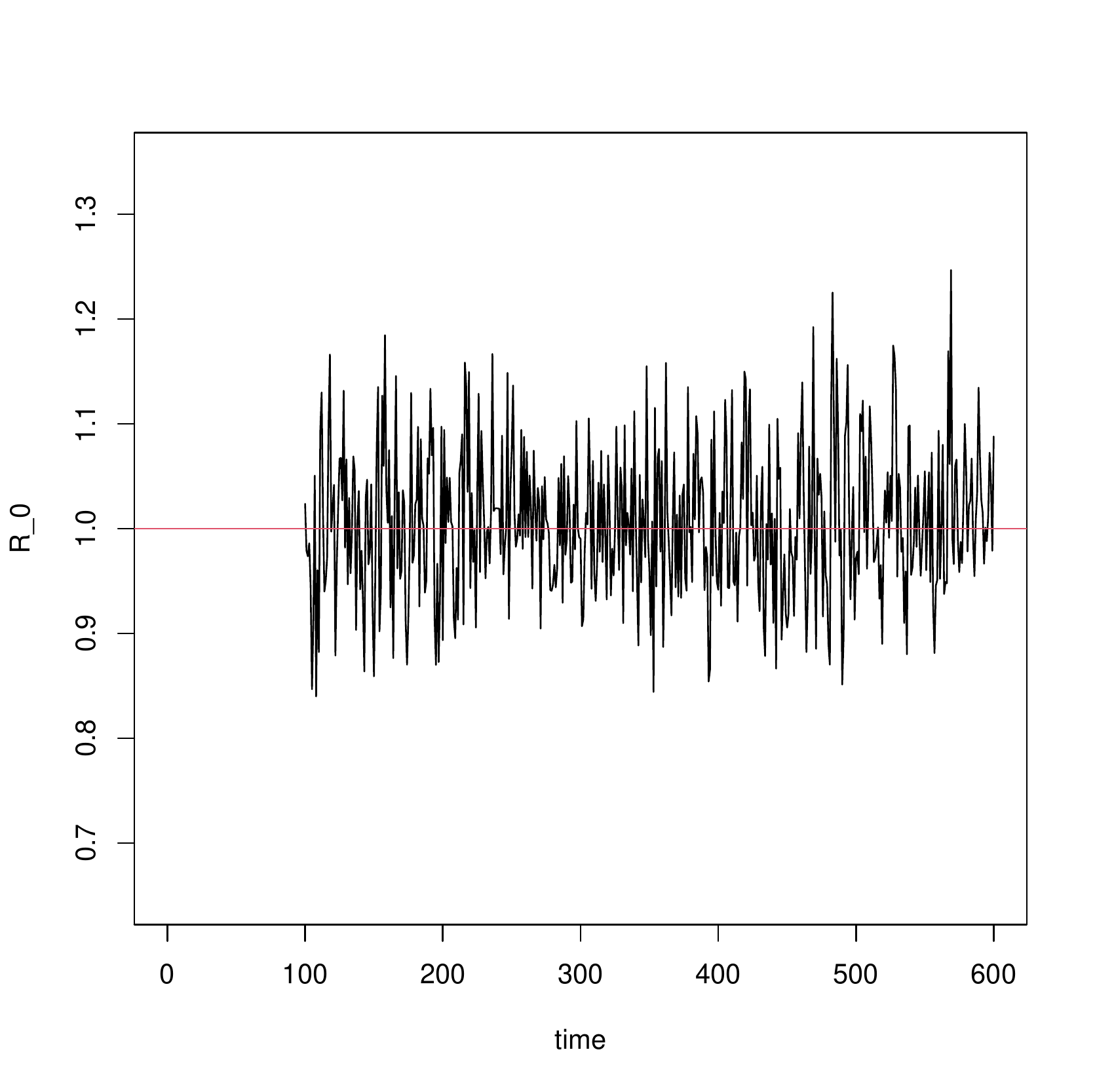}
    {\caption*{w=0.1}}
  \end{subfigure}
    \begin{subfigure}[t]{\textwidth}
  \centering
\includegraphics[height=2.2in, width=4.4in]{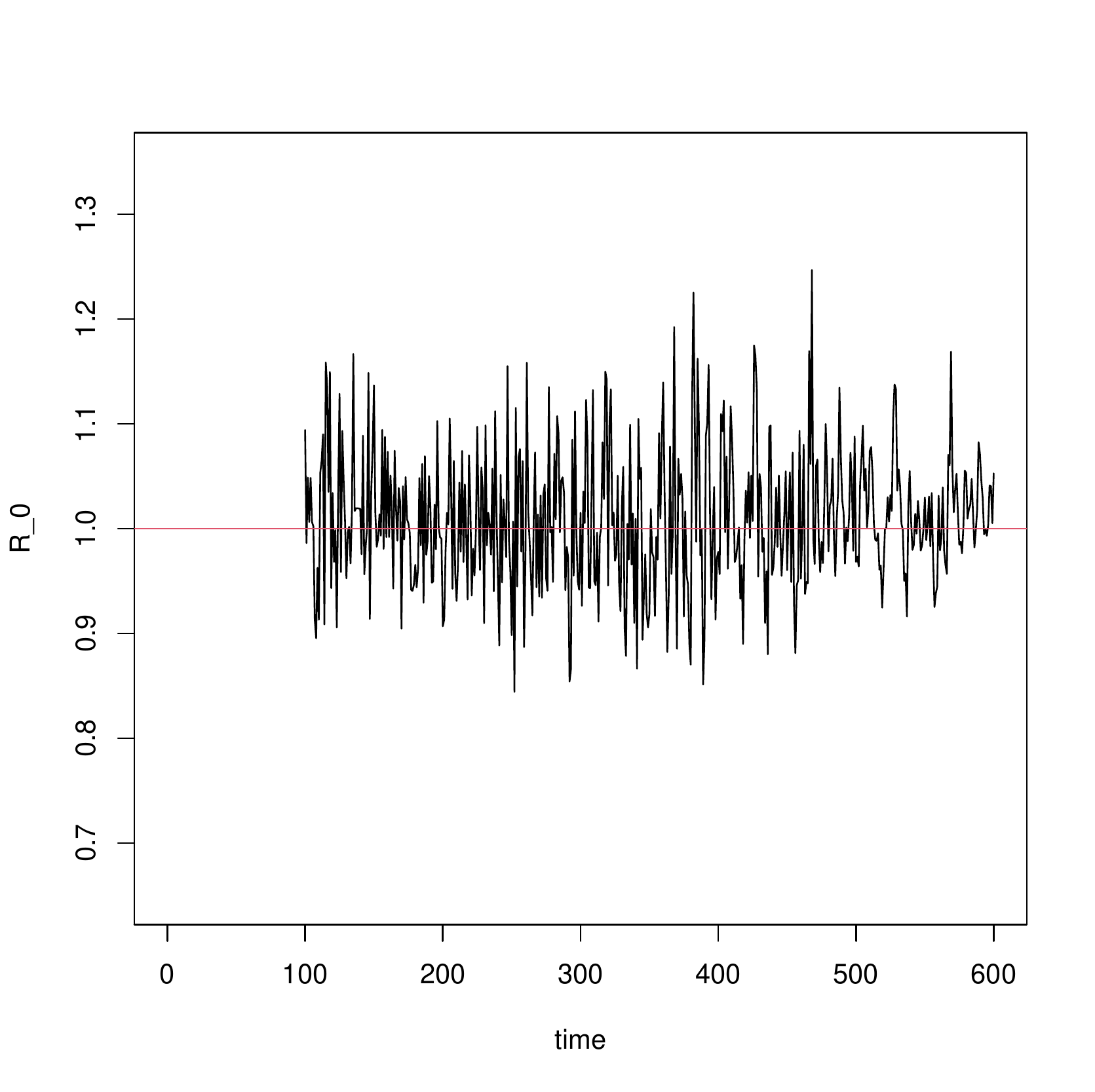}
  {\caption*{w=0.5}}
  \end{subfigure}
   \begin{subfigure}[t]{\textwidth}
   \centering
\includegraphics[height=2.2in,width=4.4in]{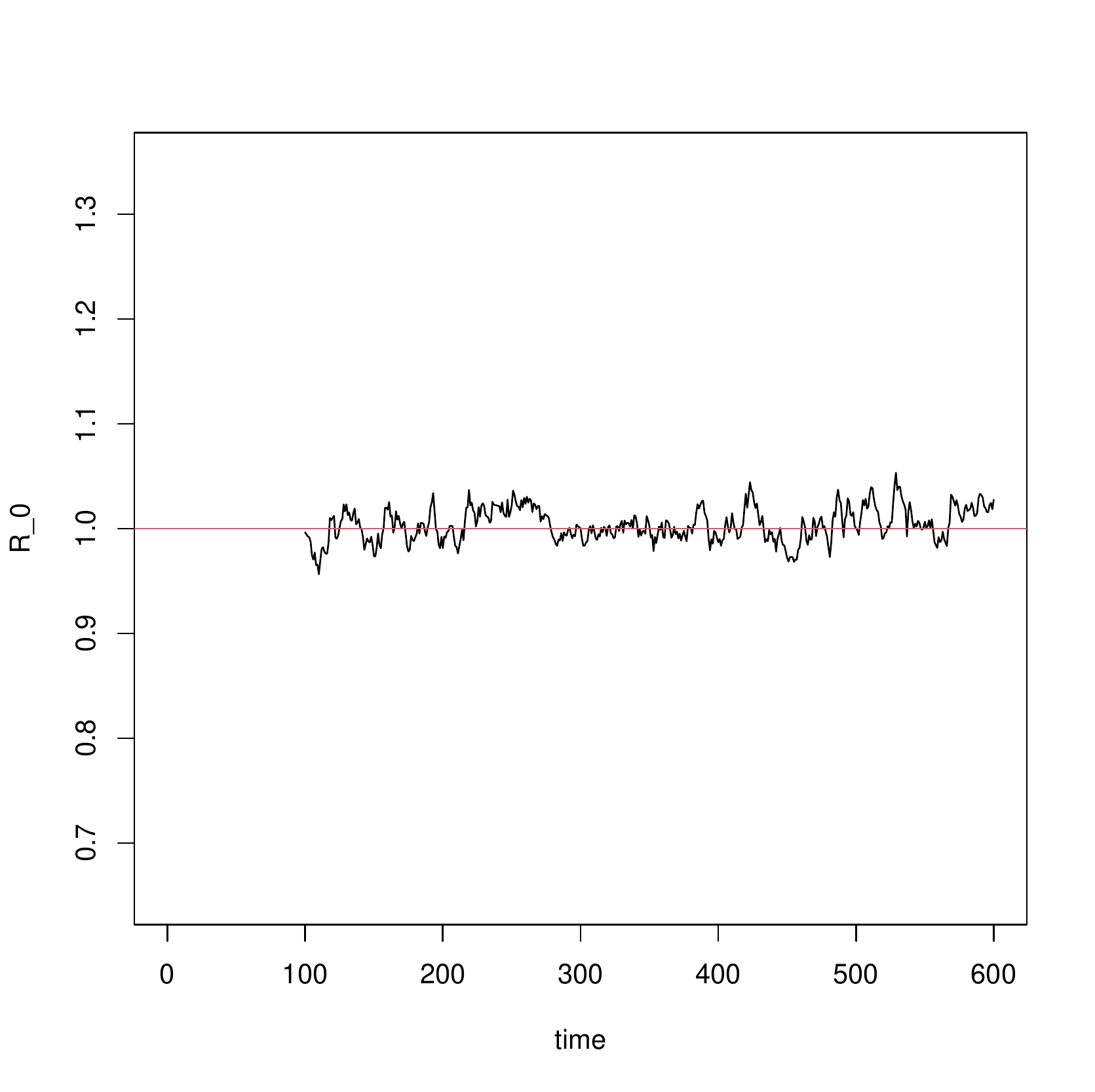}
 {\caption*{w=0.9}}
  \end{subfigure}
 Figure 3 : Estimates of Reproductive Number $R_0$, $a$  Constant
\end{figure}

\newpage

\begin{center}
\begin{figure}[h]
\centering
\includegraphics[height=2.2in]{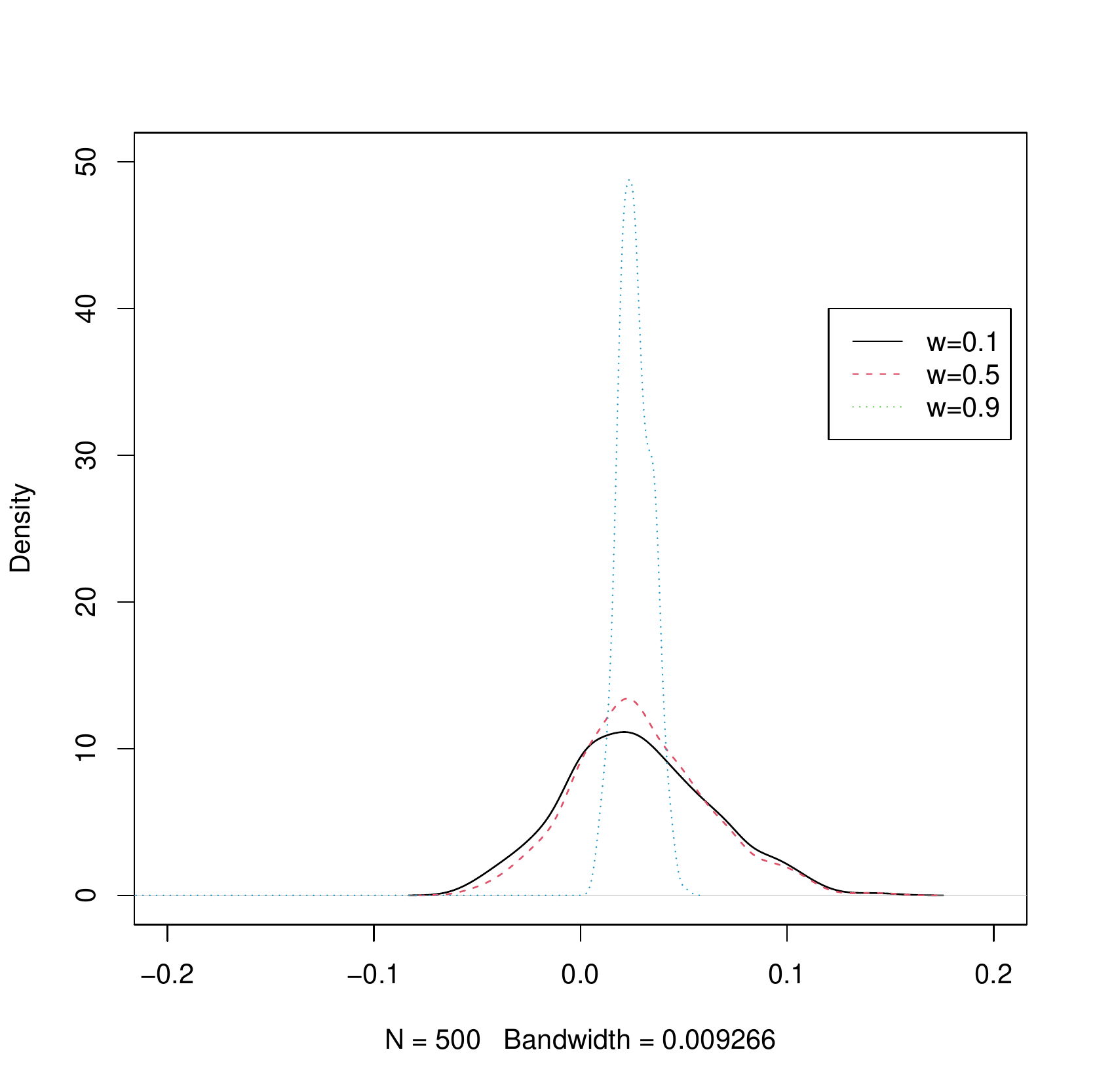}
{\caption*{Figure 4 : Deviation $\hat{a}_t - 0.2$, $a$  Constant  }}
\end{figure}
\end{center}

\begin{center}
\begin{figure}[h]
\centering
\includegraphics[height=2.2in]{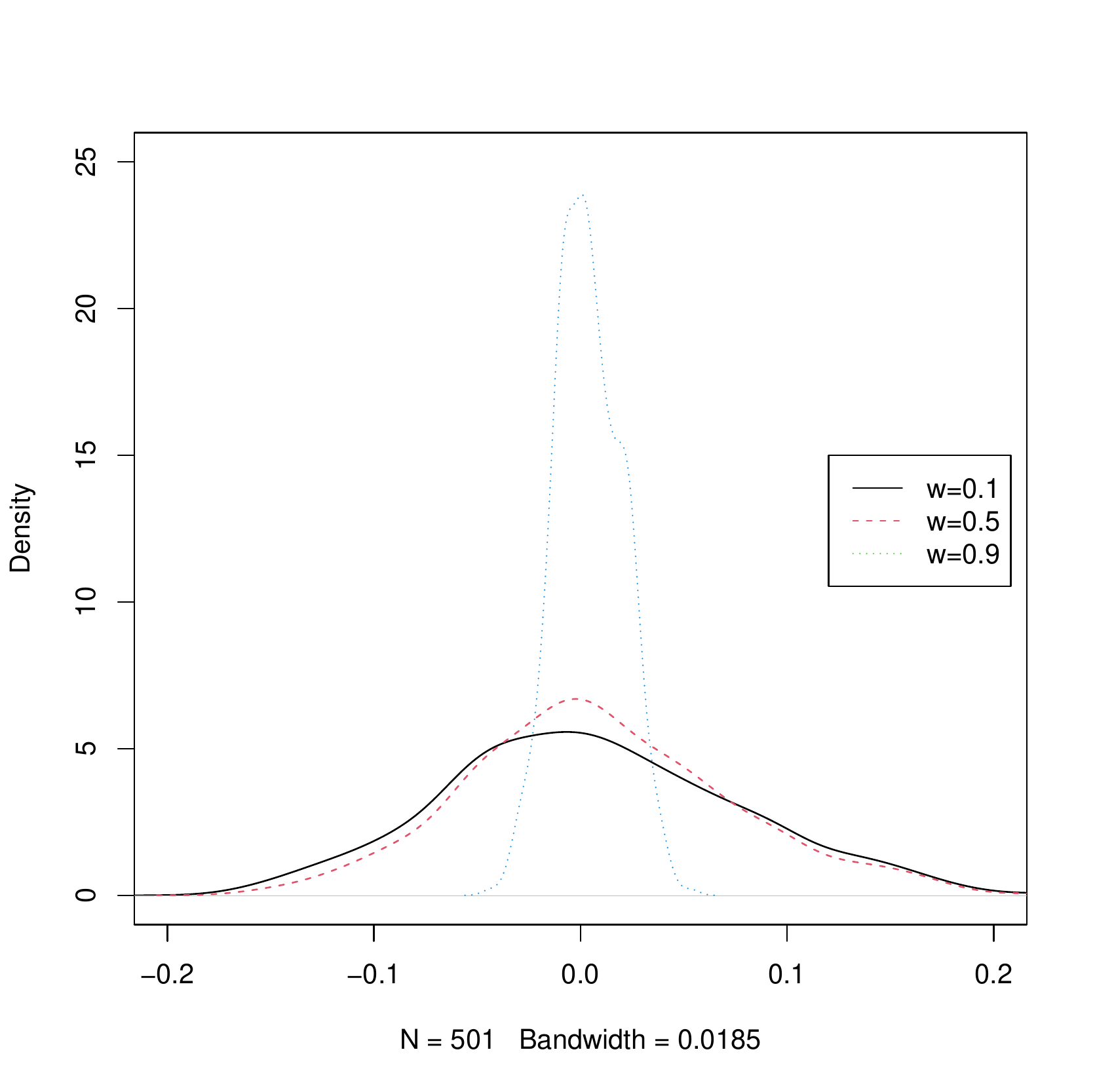}
{\caption*{Figure 5 : Deviation $\hat{R}_{0,t} - 1$, $a$  Constant  }}
\end{figure}
\end{center}

\begin{center}
\begin{figure}
\centering
\includegraphics[width=14cm, height =14cm, angle = 0]{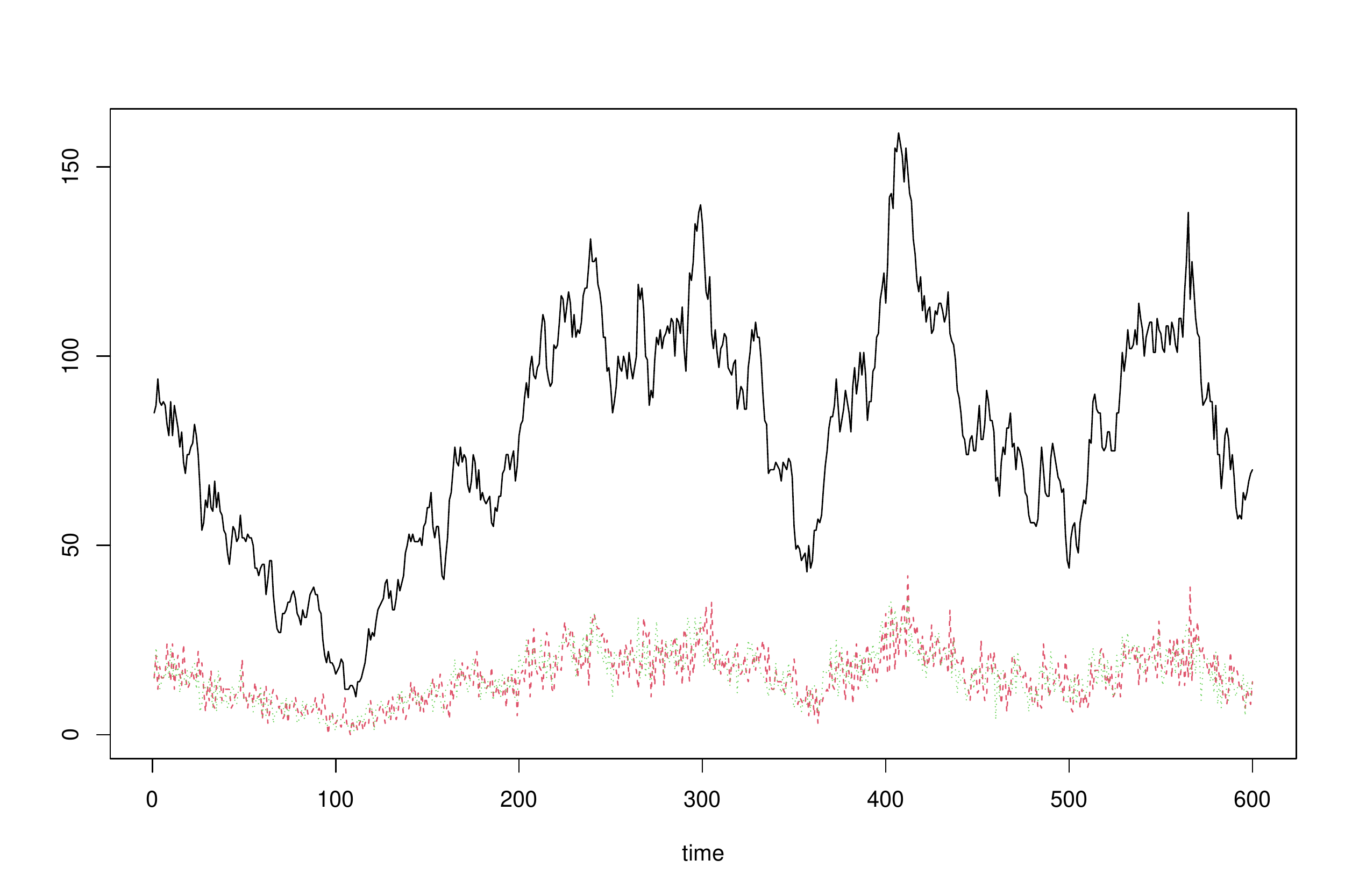}
{\caption*{Figure 6: Trajectories of Counts, Stochastic $a$  }}
\end{figure}
\end{center}

\begin{center}
\begin{figure}
\centering
\includegraphics[width=14cm,angle = 0]{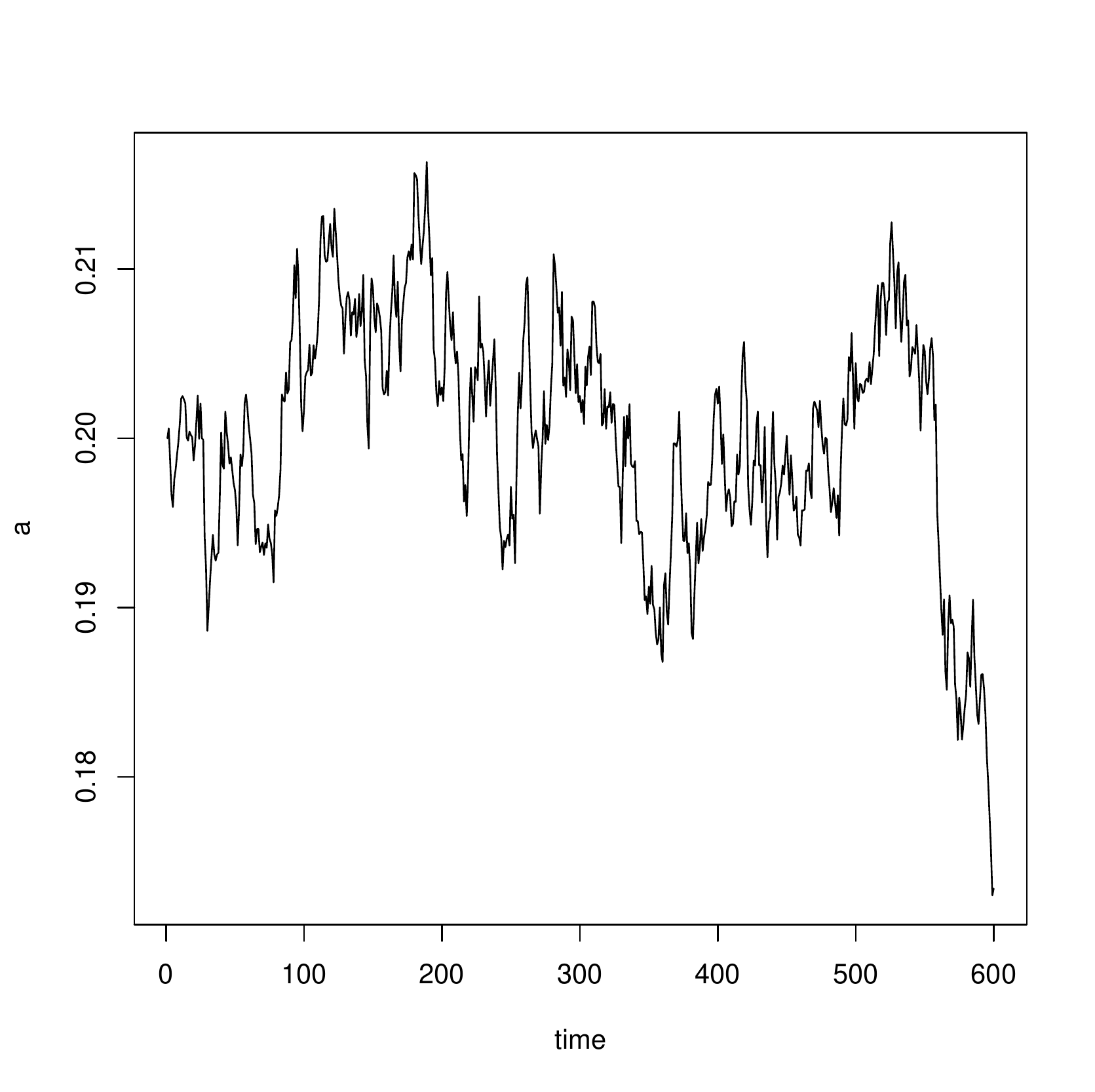}
{\caption*{Figure 7: Trajectory of Stochastic $a$  }}
\end{figure}
\end{center}

\begin{figure}
\centering
  \begin{subfigure}[t]{\textwidth}
  \centering
  \includegraphics[height=2.2in, width=4.4in]{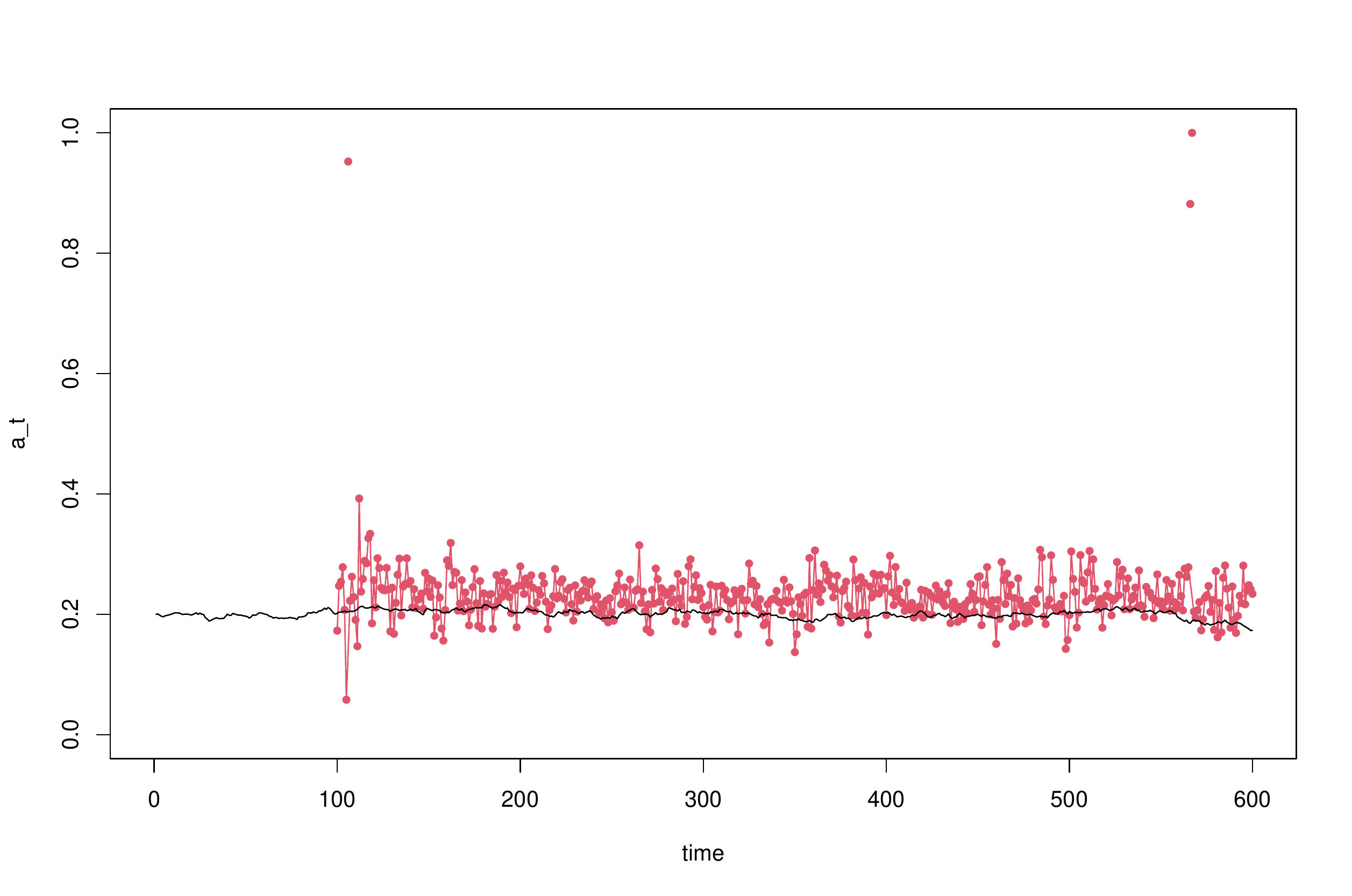}
    {\caption*{w=0.1}}
  \end{subfigure}
    \begin{subfigure}[t]{\textwidth}
  \centering
 \includegraphics[height=2.2in, width=4.4in]{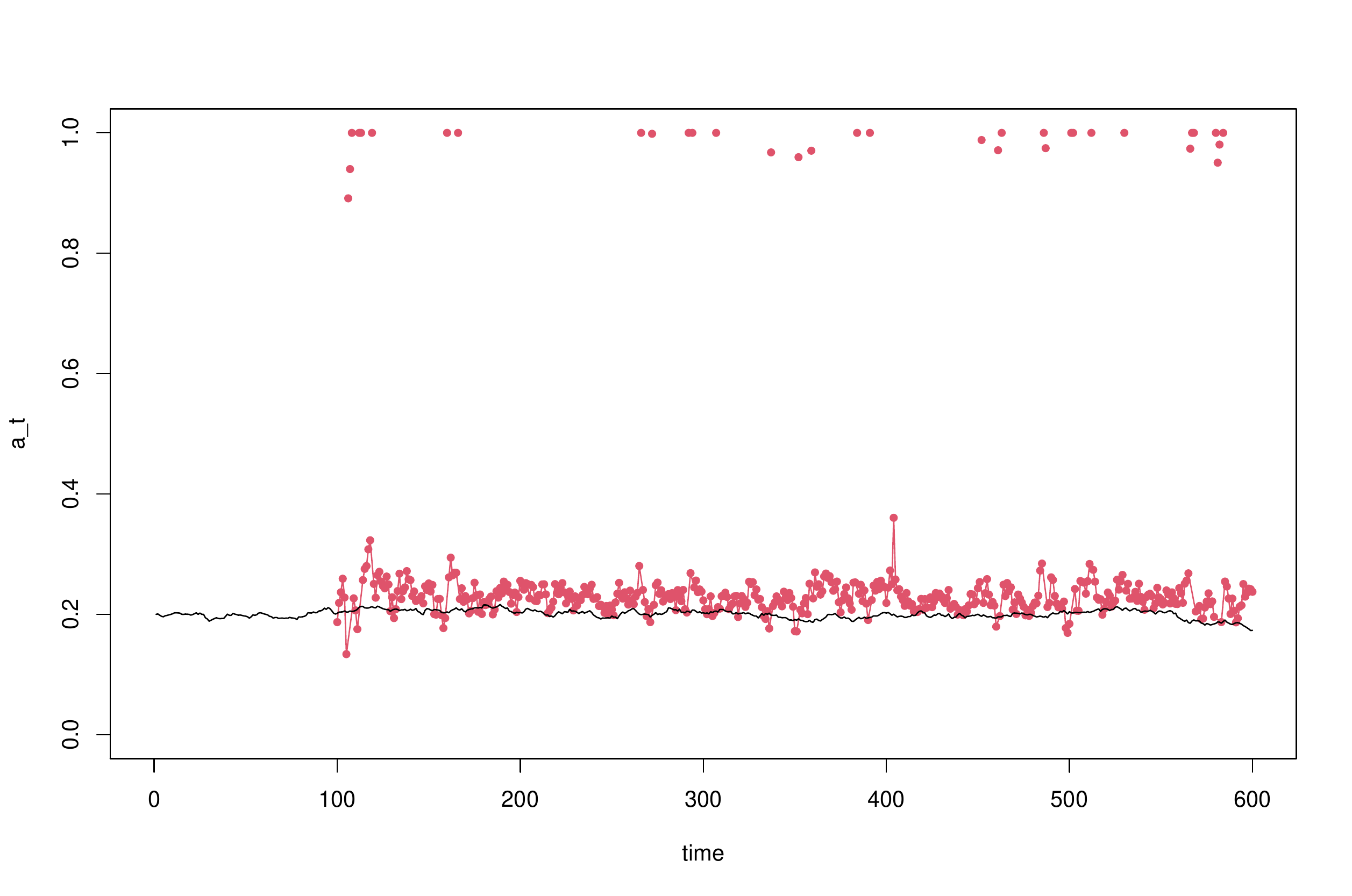}
  {\caption*{w=0.5}}
  \end{subfigure}
  \begin{subfigure}[t]{\textwidth}
  \centering
 \includegraphics[height=2.2in, width=4.4in]{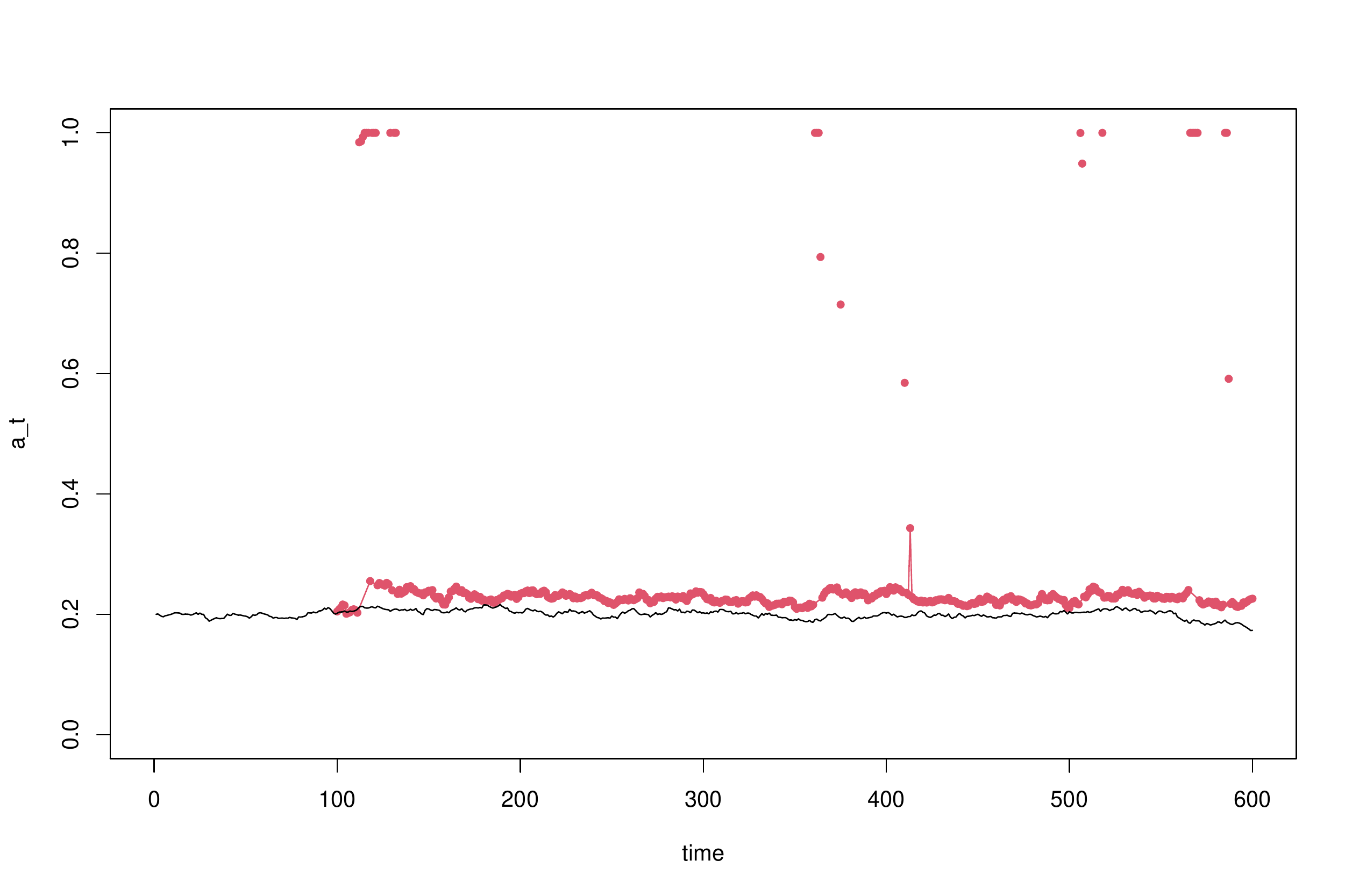}
 {\caption*{w=0.9}}
  \end{subfigure}
 Figure 8: Trajectory of Contagion Parameter Estimates, Stochastic $a$
\end{figure}

\begin{figure}
\centering
  \begin{subfigure}[t]{\textwidth}
  \centering
\includegraphics[height=2.2in, width=4.4in]{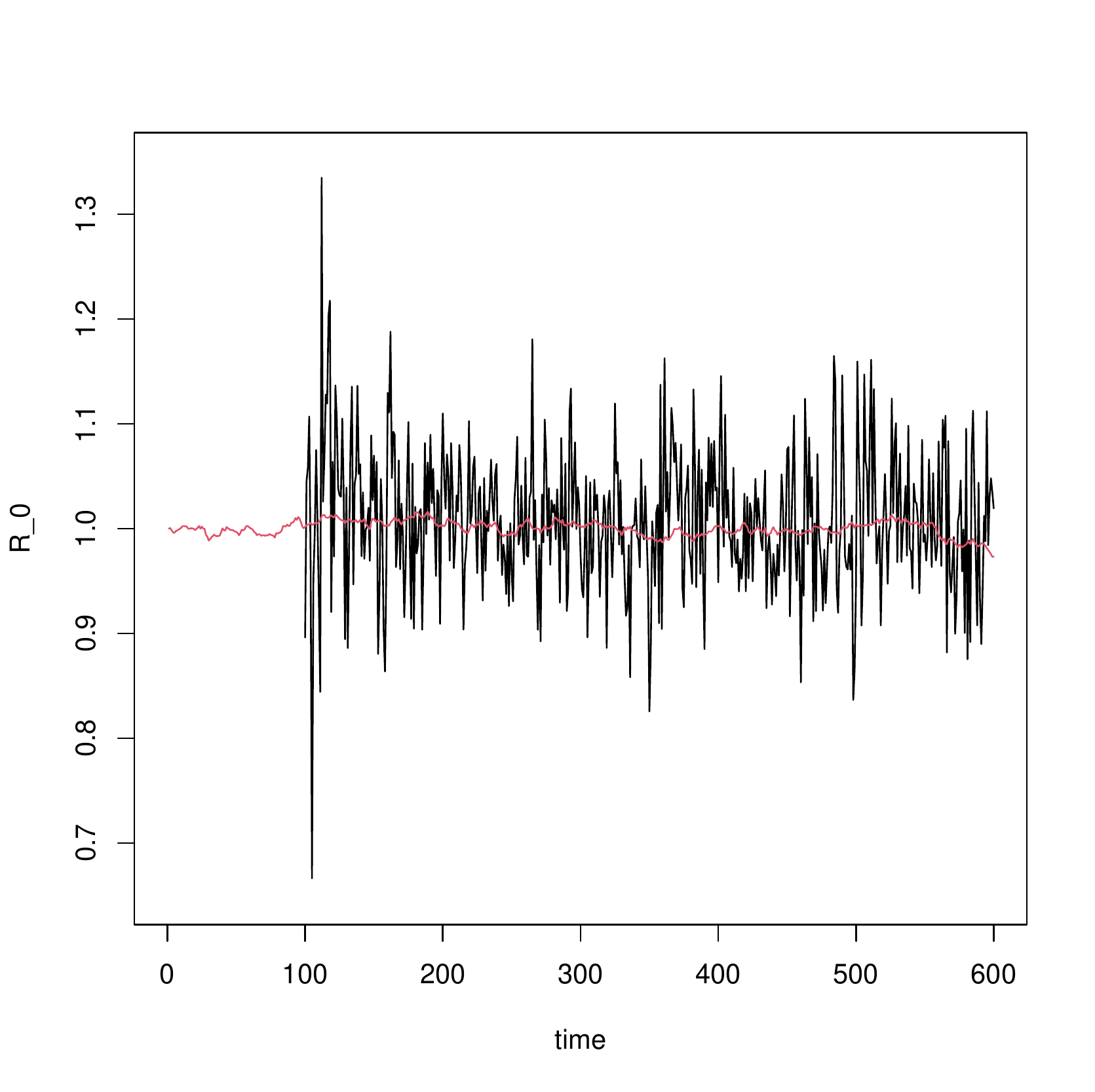}
    {\caption*{w=0.1}}
  \end{subfigure}
    \begin{subfigure}[t]{\textwidth}
  \centering
\includegraphics[height=2.2in, width=4.4in]{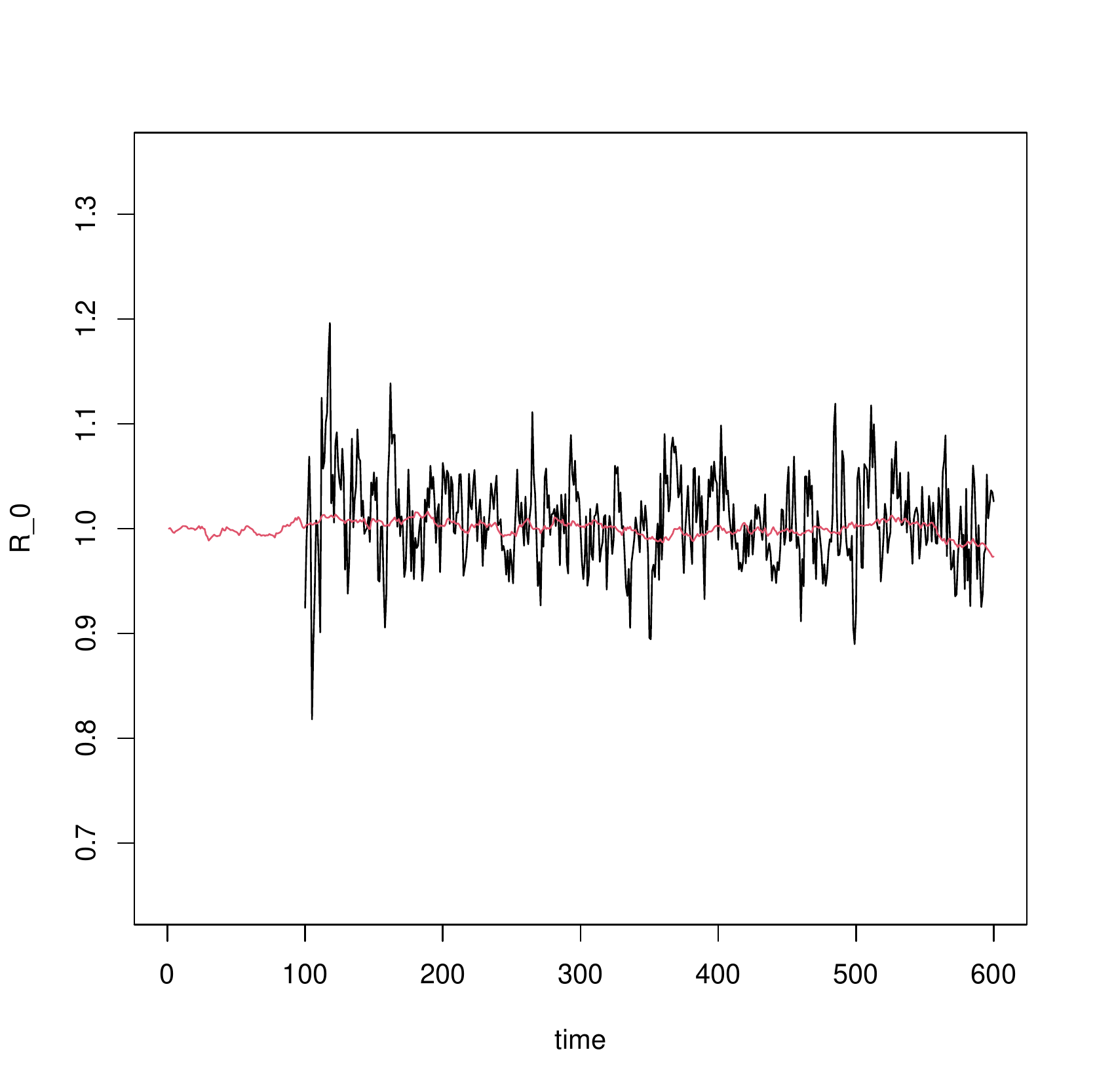}
  {\caption*{w=0.5}}
  \end{subfigure}
   \begin{subfigure}[t]{\textwidth}
   \centering
\includegraphics[height=2.2in, width=4.4in]{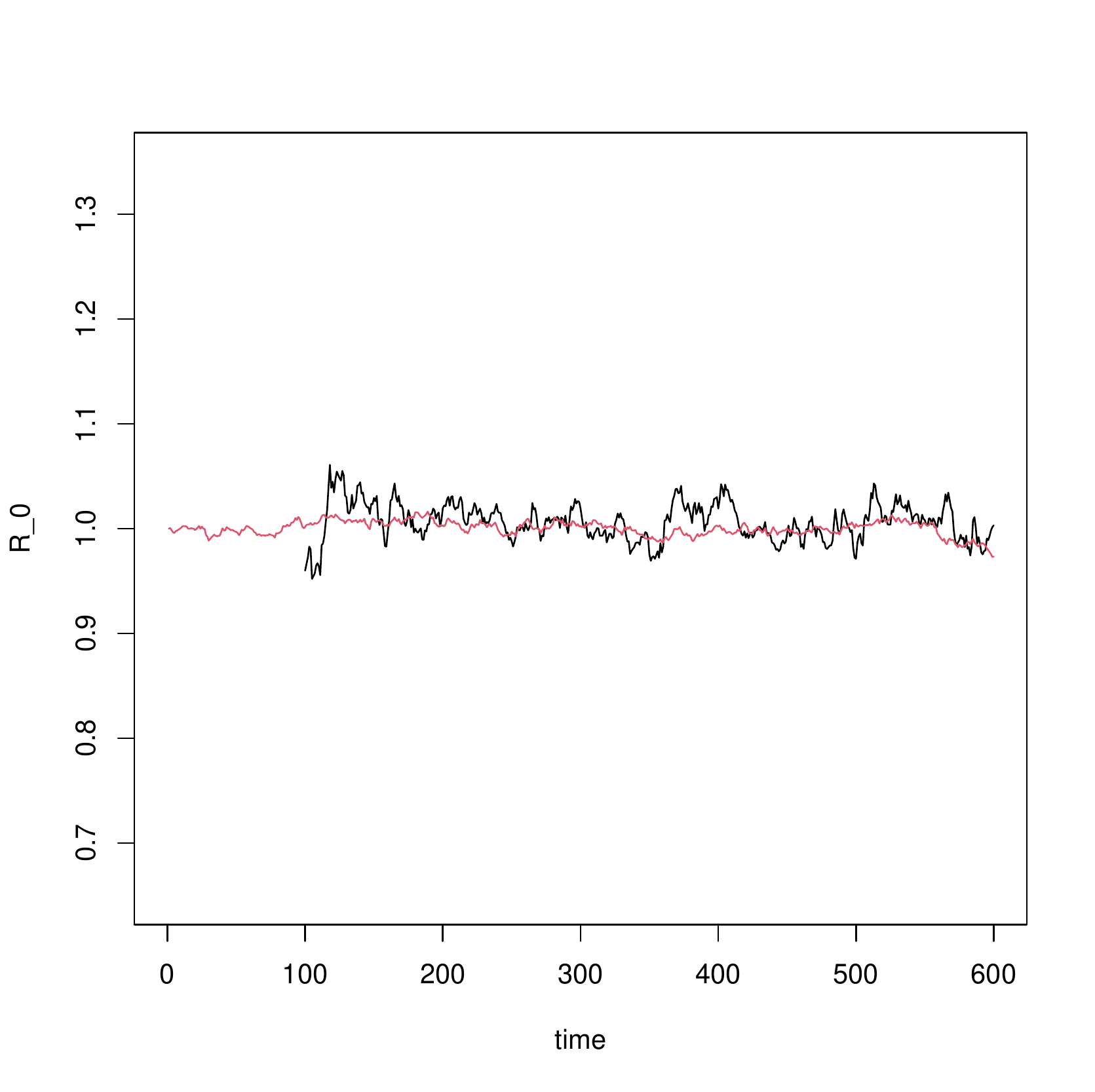}
 {\caption*{w=0.9}}
  \end{subfigure}
 Figure 9: Estimates of Reproductive Number $R_0$, Stochastic $a$
\end{figure}

\begin{figure}
\centering
  \begin{subfigure}[t]{\textwidth}
  \centering
\includegraphics[width =2.2in, height=2.2in]{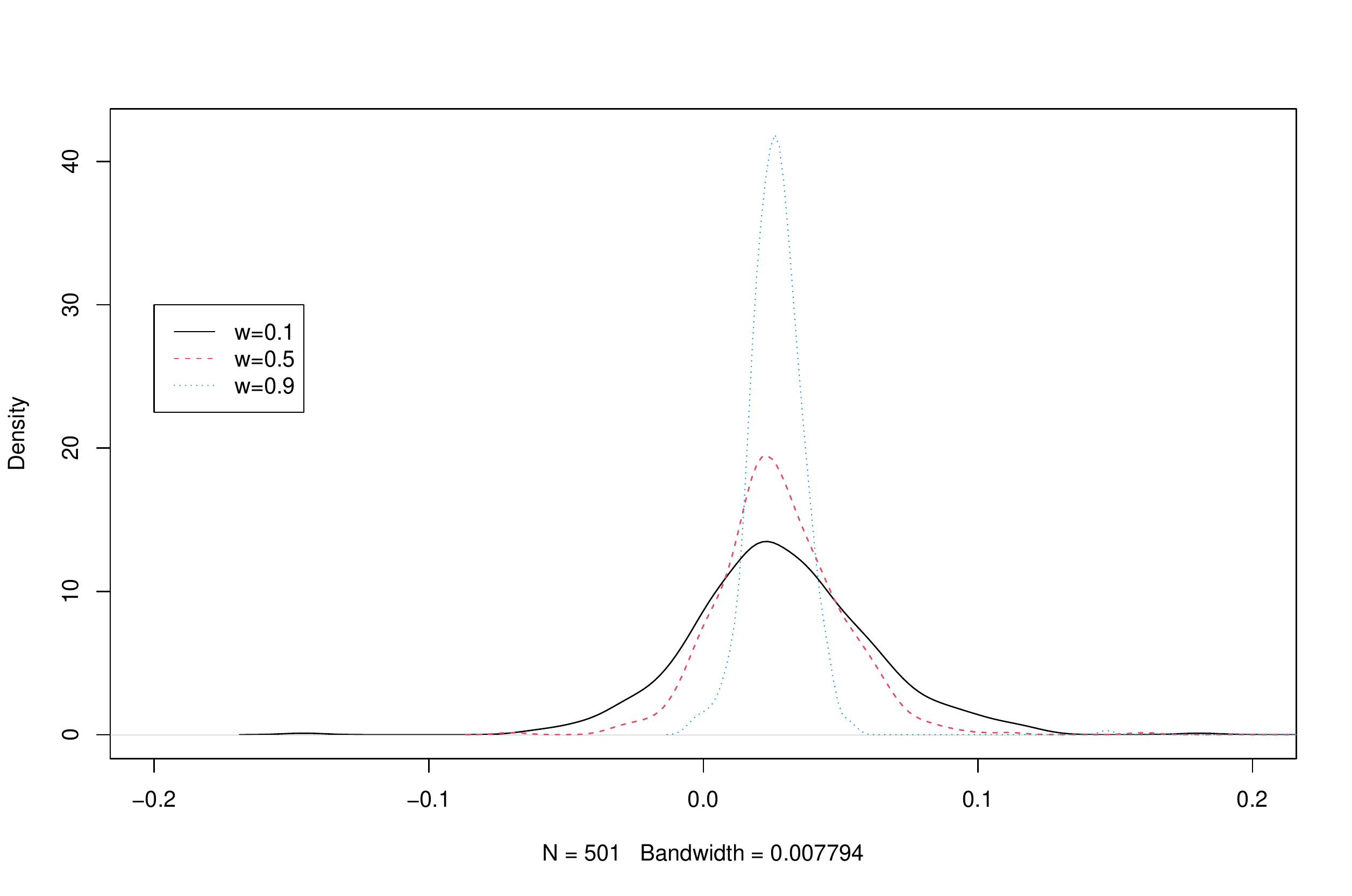}
    {\caption*{Figure 10: Deviation $\hat{a}_t -a_t$,  Stochastic $a$}}
  \end{subfigure}
    \begin{subfigure}[t]{\textwidth}
  \centering
\includegraphics[width =2.2in, height=2.2in]{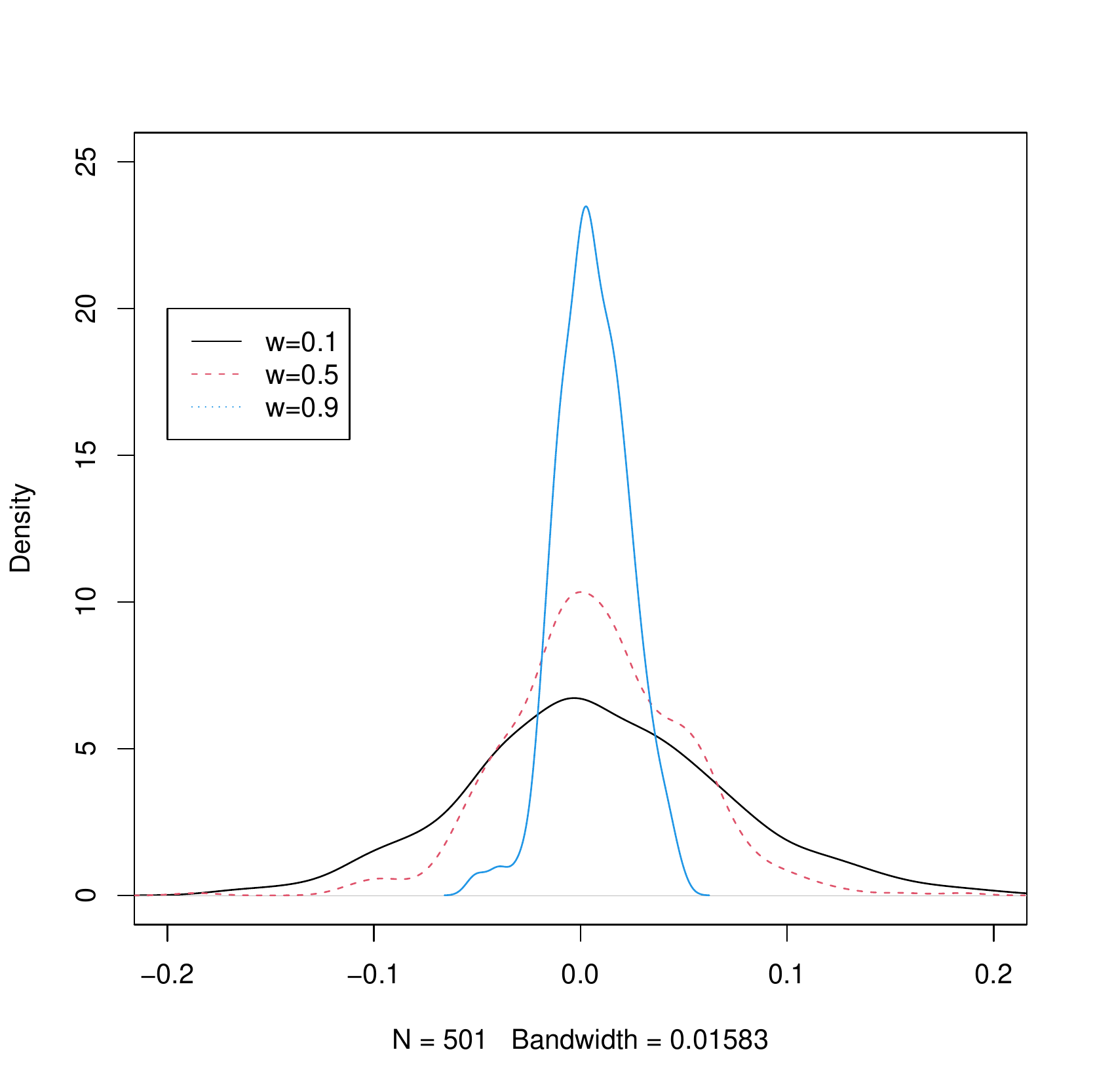}
  {\caption*{Figure 11: Deviation $\hat{R}_{0,t} - R_{0,t}$, Stochastic $a$}}
  \end{subfigure}
\end{figure}

\begin{figure}
\centering
  \begin{subfigure}[t]{\textwidth}
  \centering
  \includegraphics[height=2.2in]{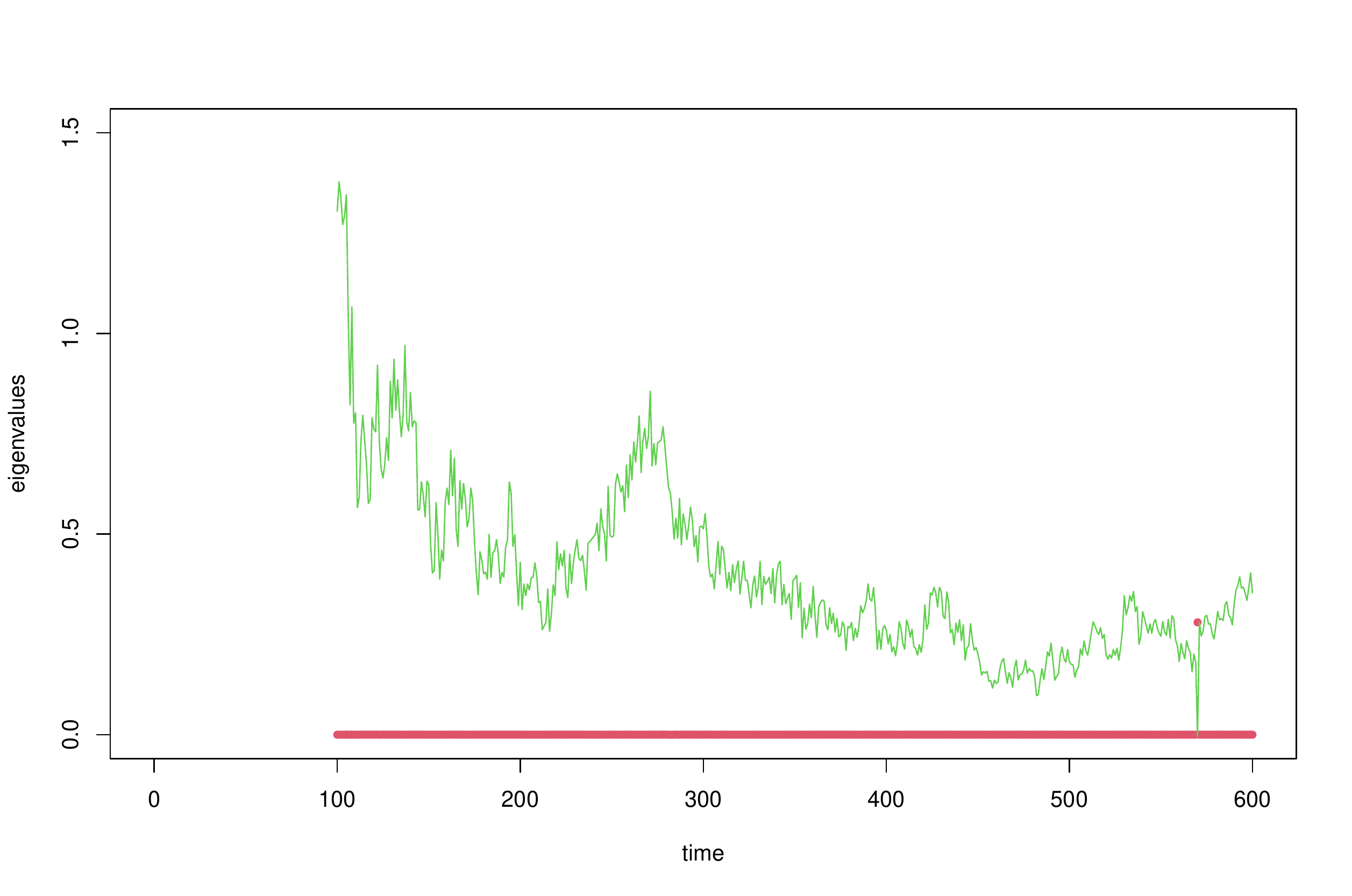}
    {\caption*{w=0.1}}
  \end{subfigure}
    \begin{subfigure}[t]{\textwidth}
  \centering
 \includegraphics[height=2.2in]{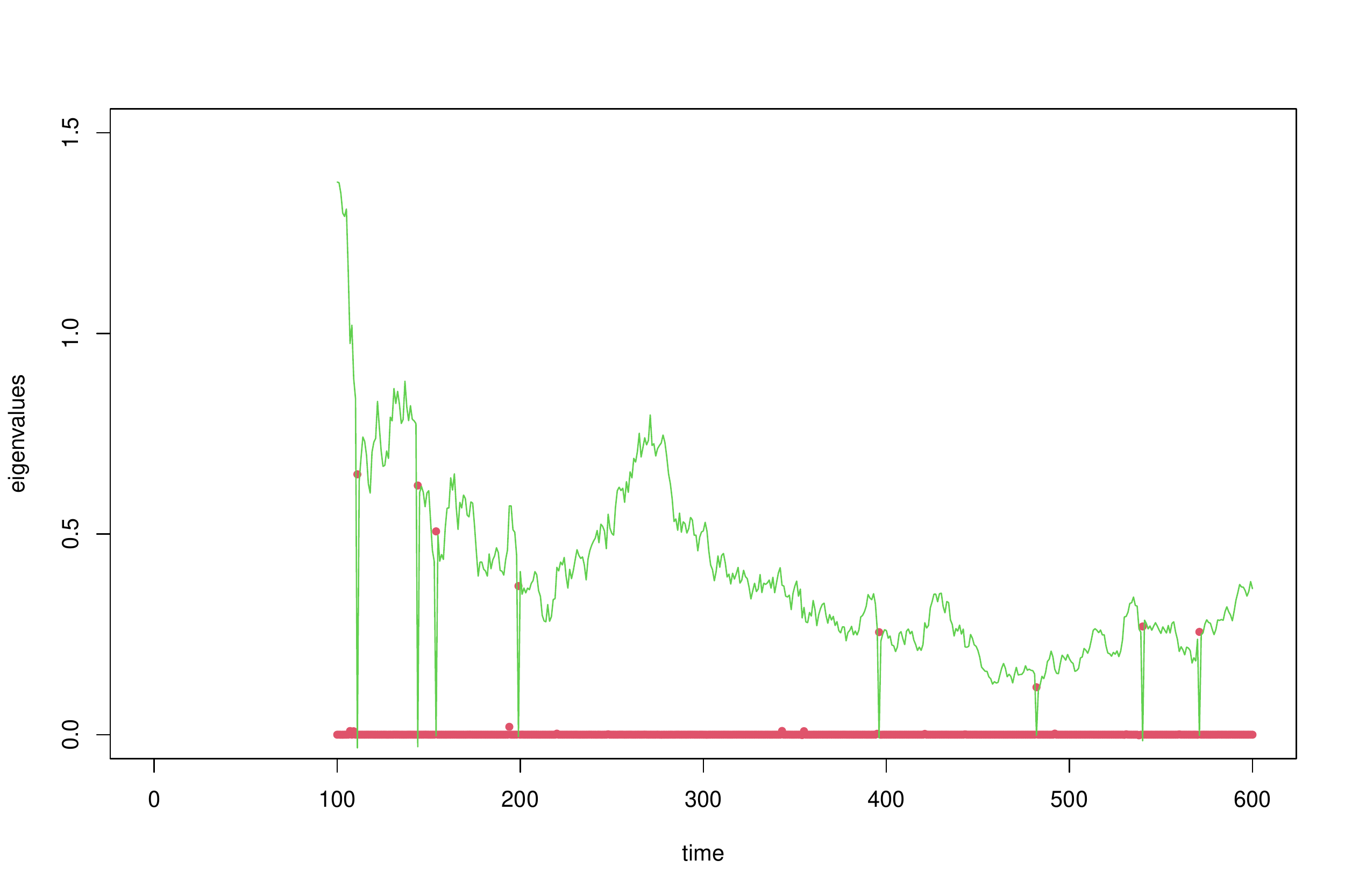}
  {\caption*{w=0.5}}
  \end{subfigure}
   \begin{subfigure}[t]{\textwidth}
   \centering
 \includegraphics[height=2.2in]{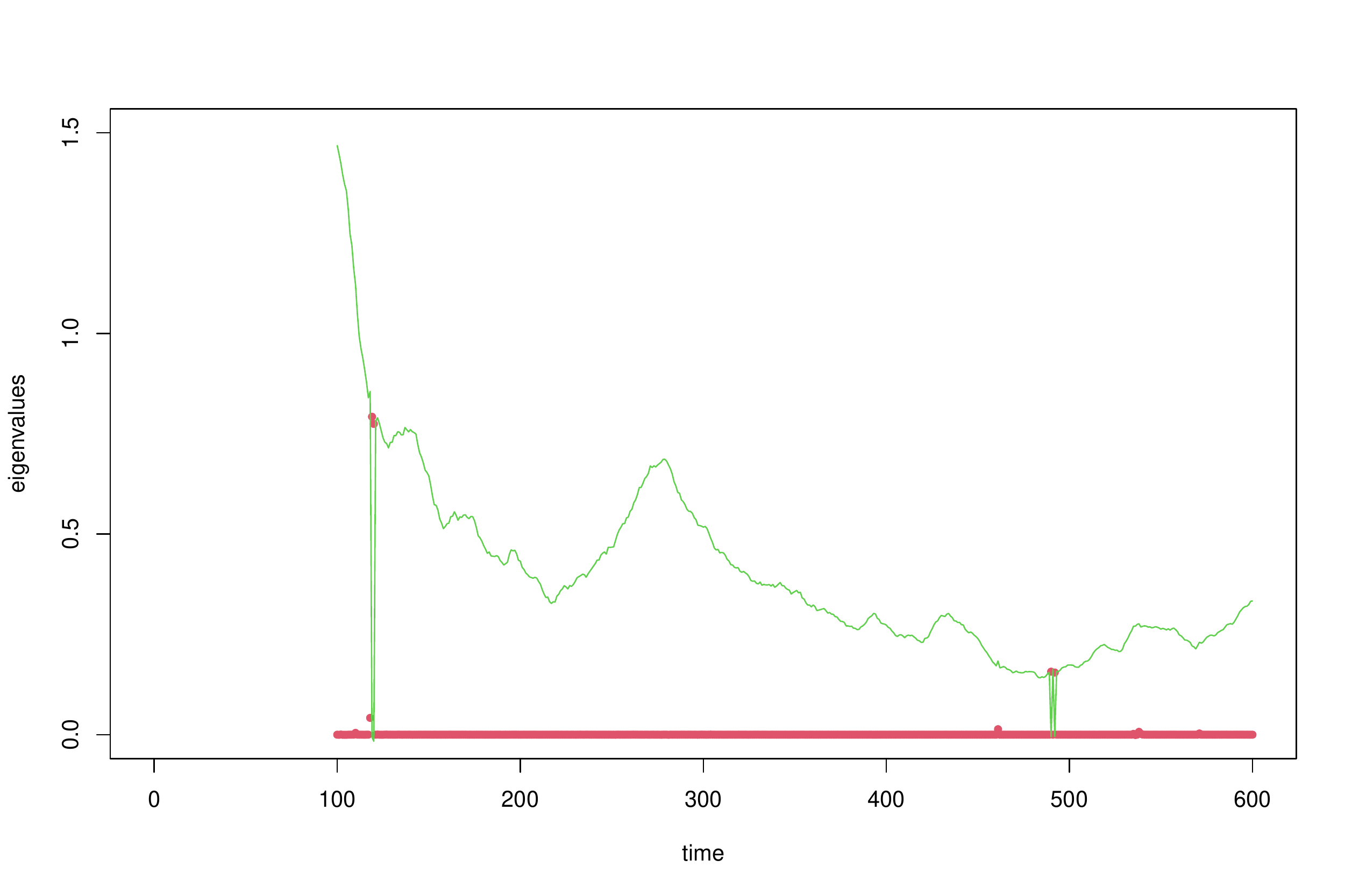}
 {\caption*{w=0.9}}
  \end{subfigure}
 Figure 12: Eigenvalues of Estimated Information Matrix: Constant $a$
\end{figure}

\begin{figure}
\centering
  \begin{subfigure}[t]{\textwidth}
  \centering
  \includegraphics[height=2.2in]{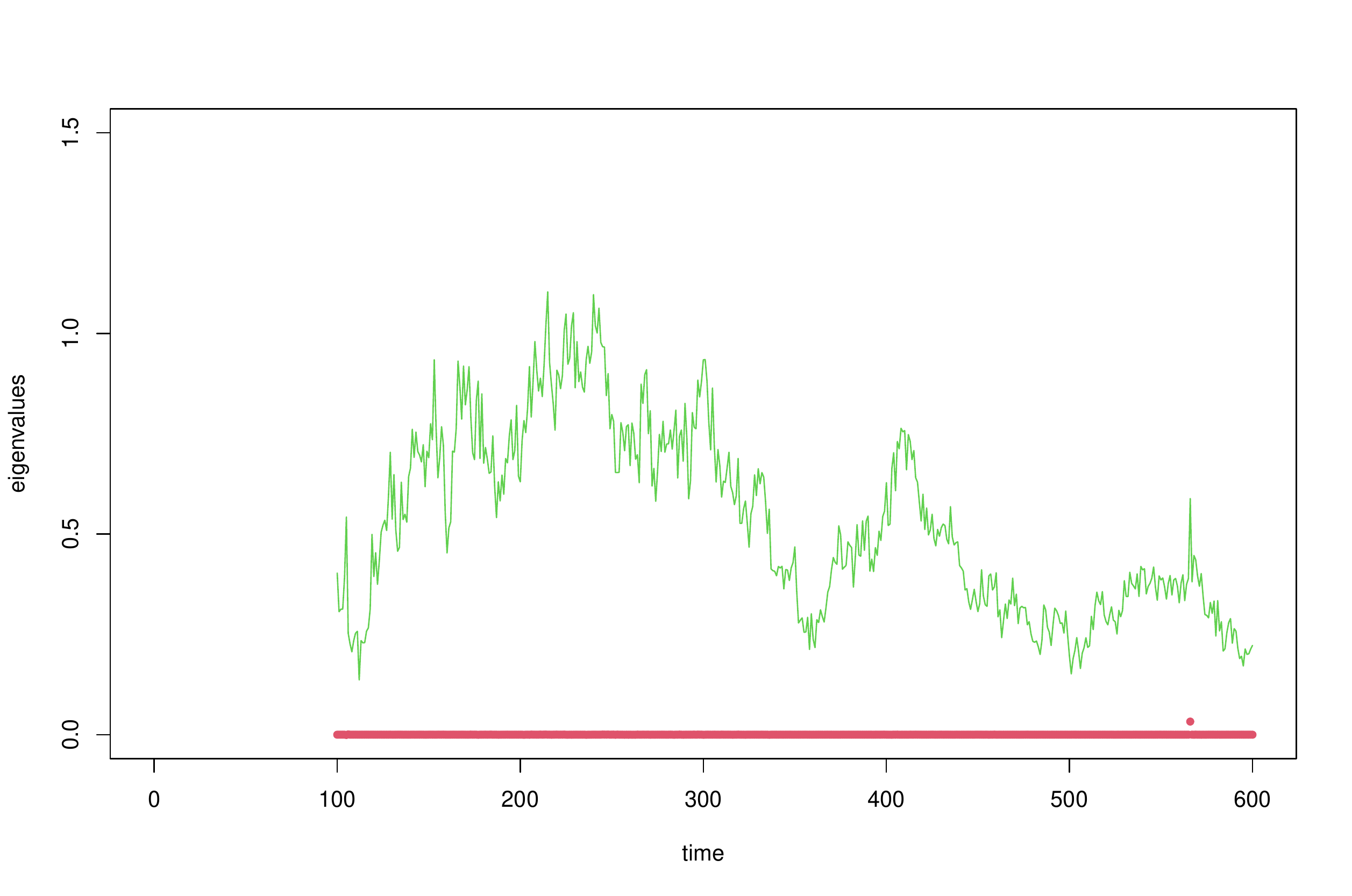}
    {\caption*{w=0.1}}
  \end{subfigure}
    \begin{subfigure}[t]{\textwidth}
  \centering
 \includegraphics[height=2.2in]{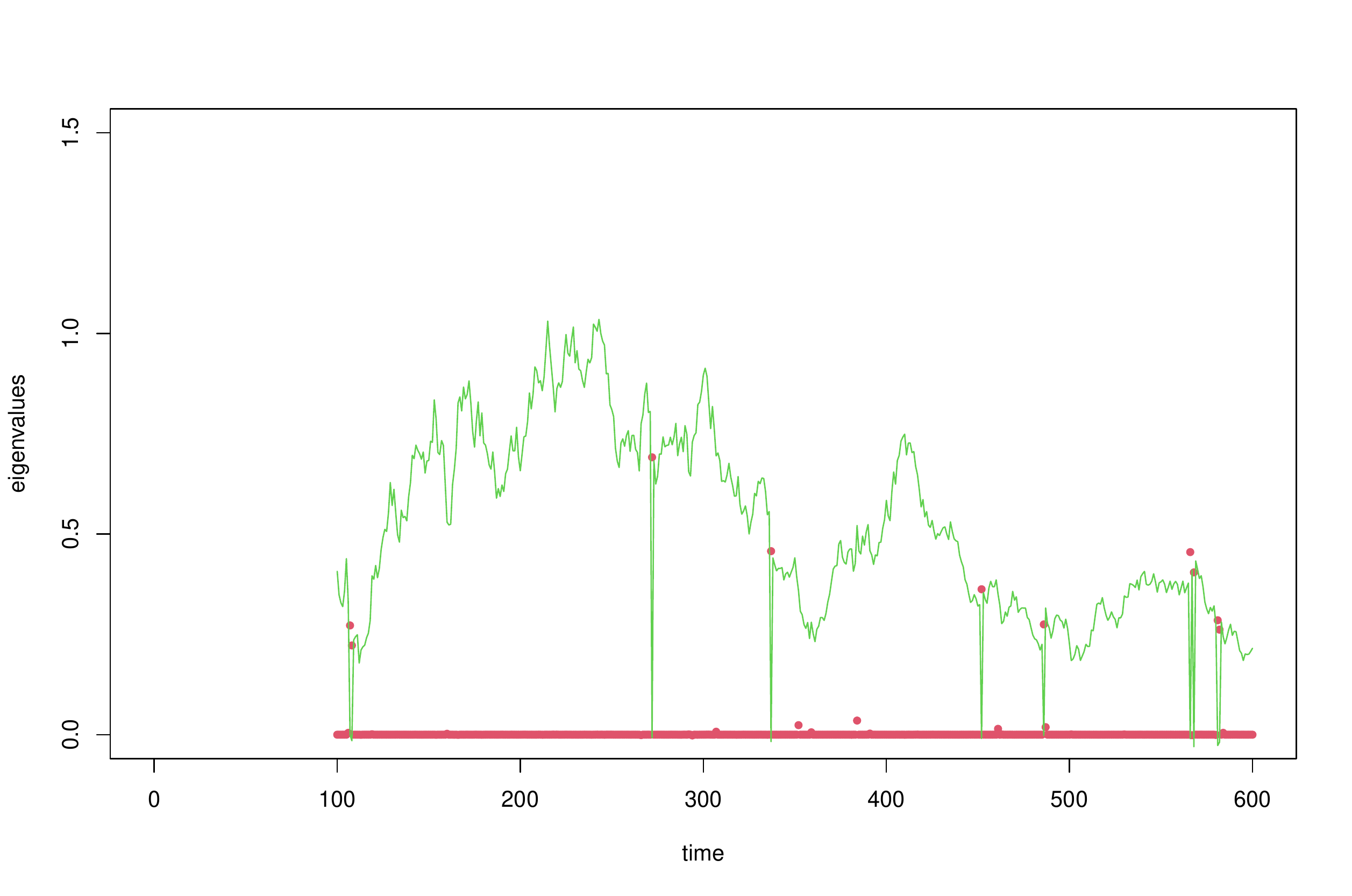}
  {\caption*{w=0.5}}
  \end{subfigure}
   \begin{subfigure}[t]{\textwidth}
   \centering
 \includegraphics[height=2.2in]{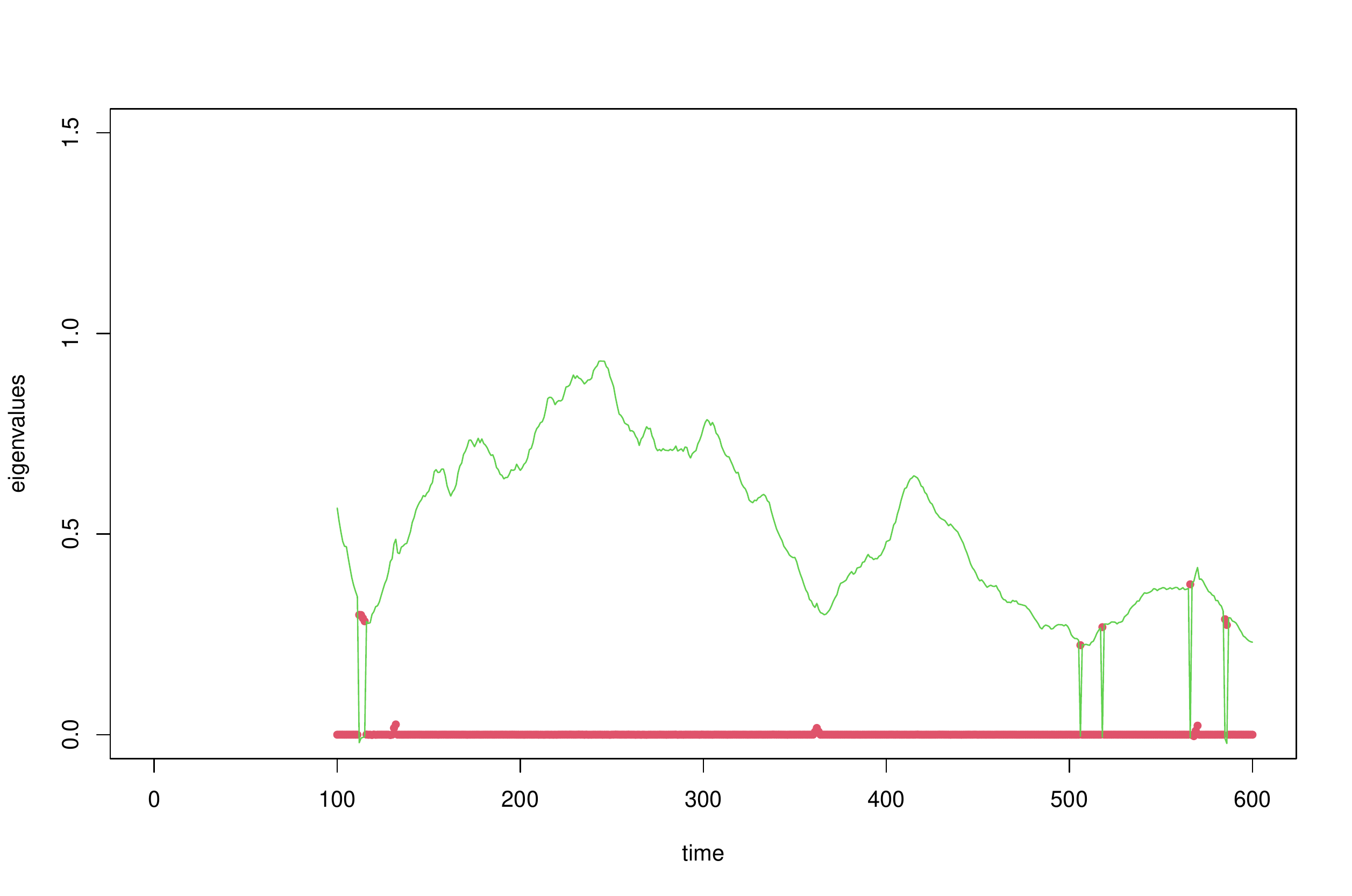}
 {\caption*{w=0.9}}
  \end{subfigure}
  \medskip
 Figure 13: Eigenvalues of Estimated Information Matrix: Stochastic $a$
\end{figure}

\begin{figure}
\centering
  \begin{subfigure}[t]{\textwidth}
  \centering
  \includegraphics[height=2.2in]{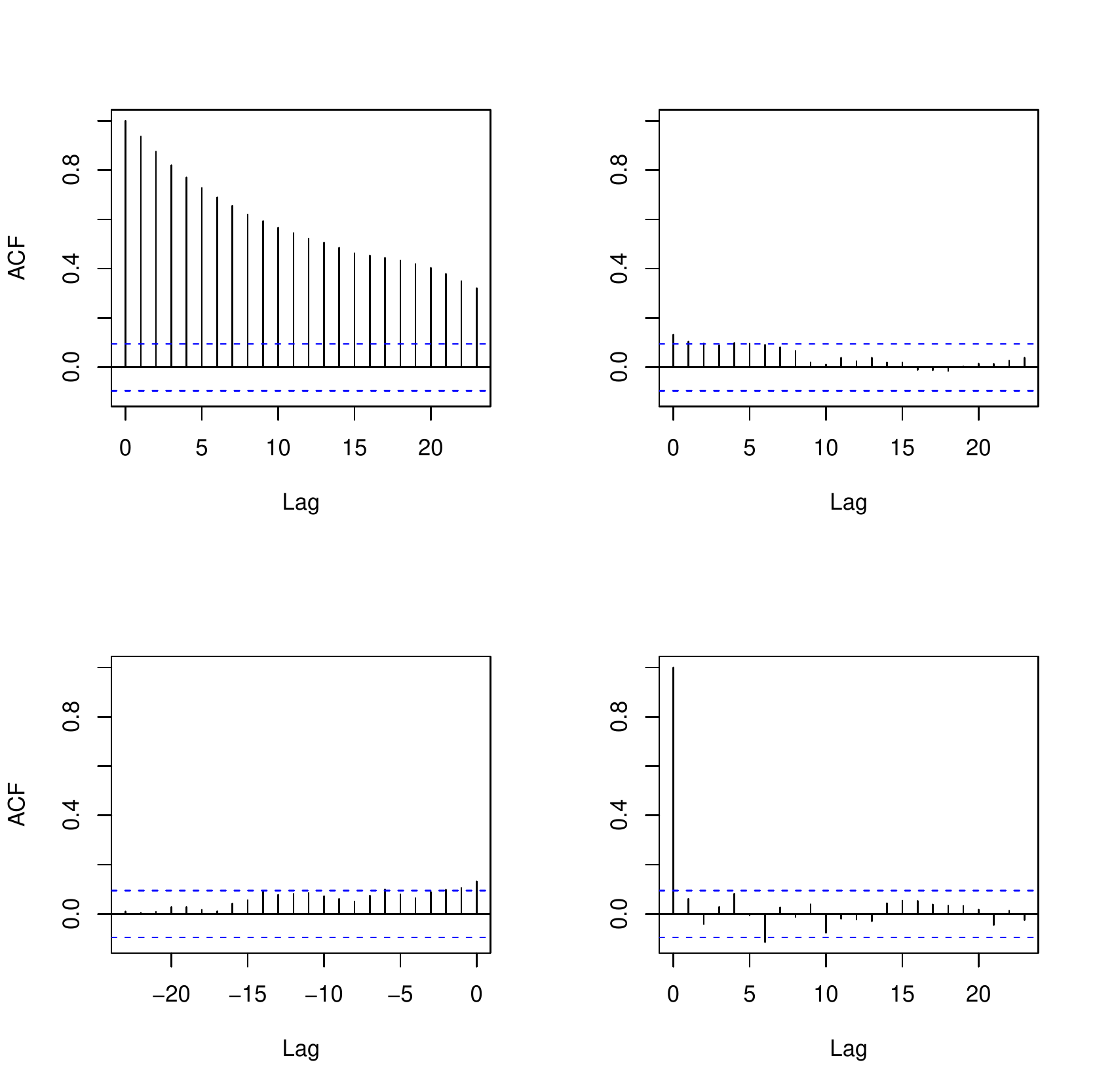}
    {\caption*{w=0.1}}
  \end{subfigure}
    \begin{subfigure}[t]{\textwidth}
  \centering
 \includegraphics[height=2.2in]{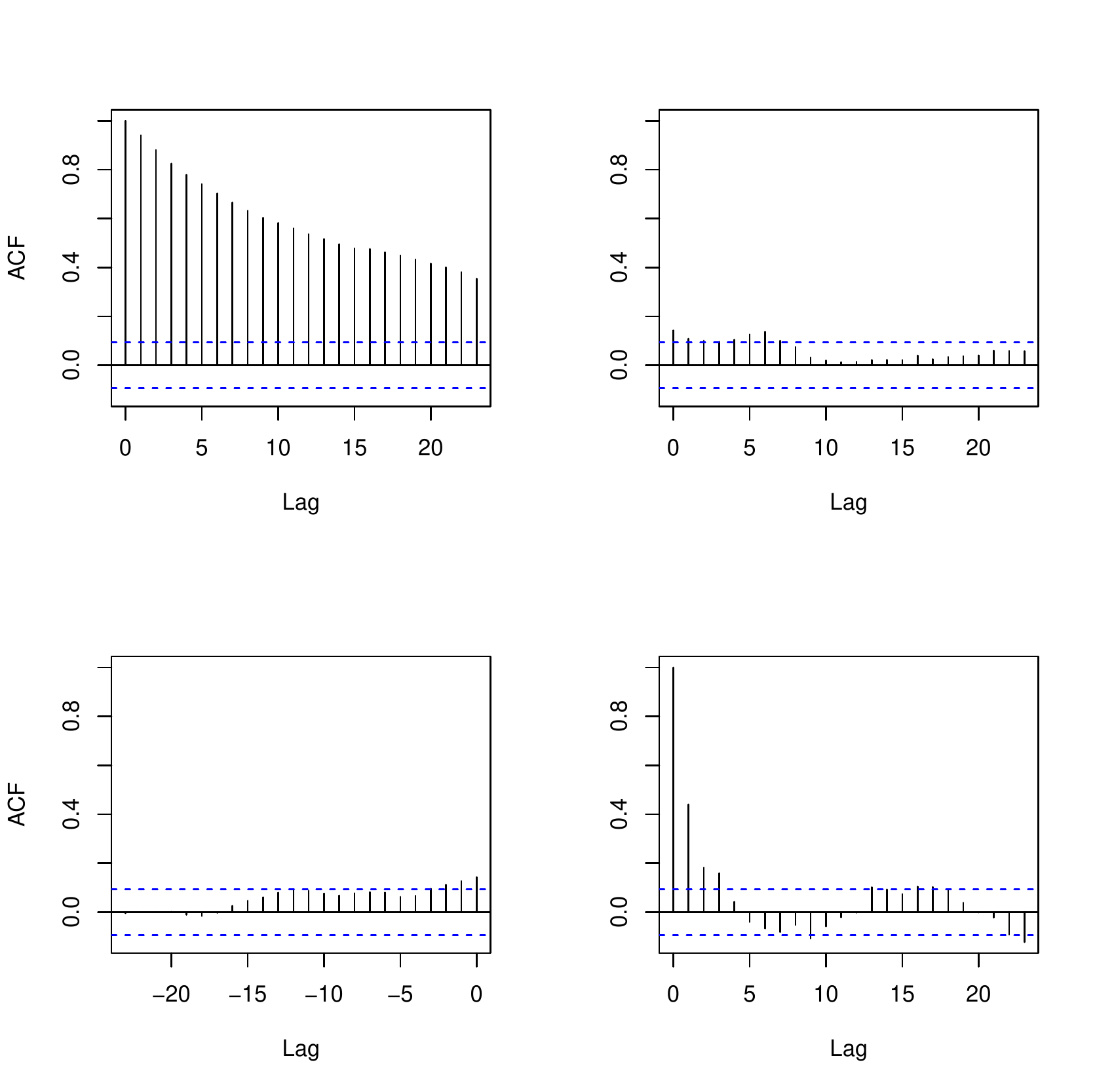}
  {\caption*{w=0.5}}
  \end{subfigure}
   \begin{subfigure}[t]{\textwidth}
   \centering
 \includegraphics[height=2.2in]{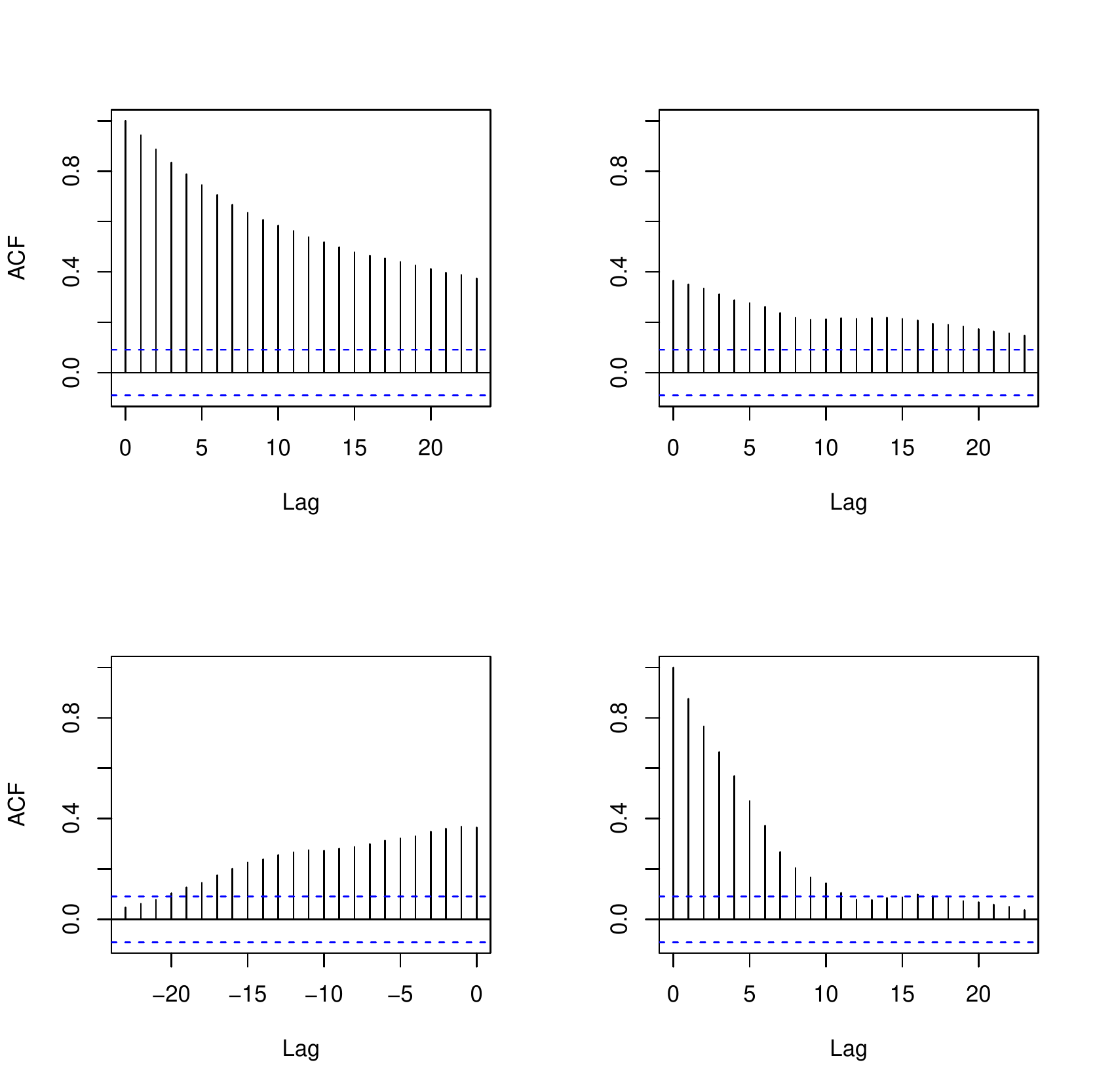}
 {\caption*{w=0.9}}
  \end{subfigure}
  \medskip
 Figure 14: Joint ACF of $a_t$ and $\hat{a}_t$
\end{figure}

\begin{figure}
\centering
  \begin{subfigure}[t]{\textwidth}
  \centering
  \includegraphics[height=2.2in]{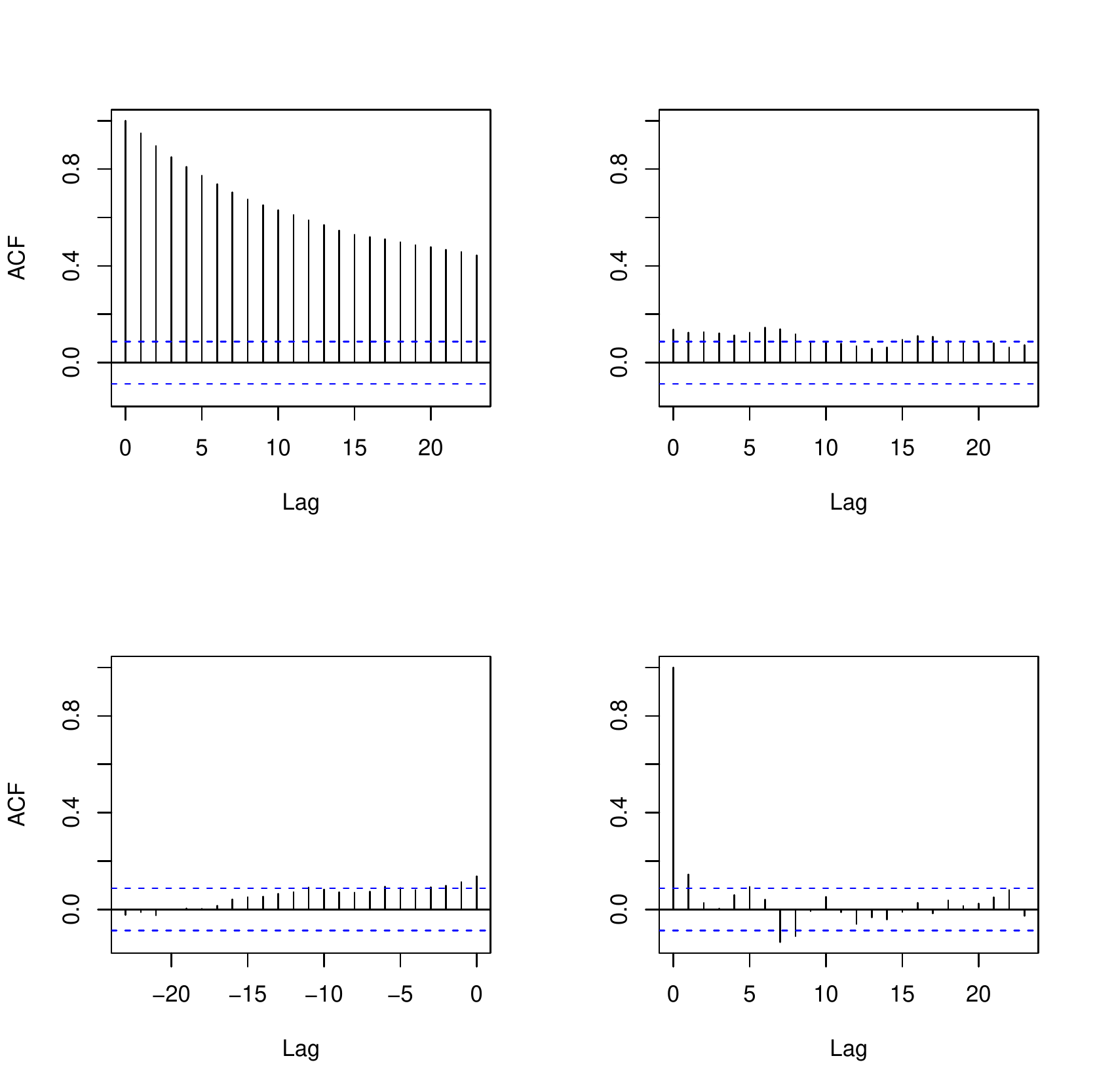}
    {\caption*{w=0.1}}
  \end{subfigure}
    \begin{subfigure}[t]{\textwidth}
  \centering
 \includegraphics[height=2.2in]{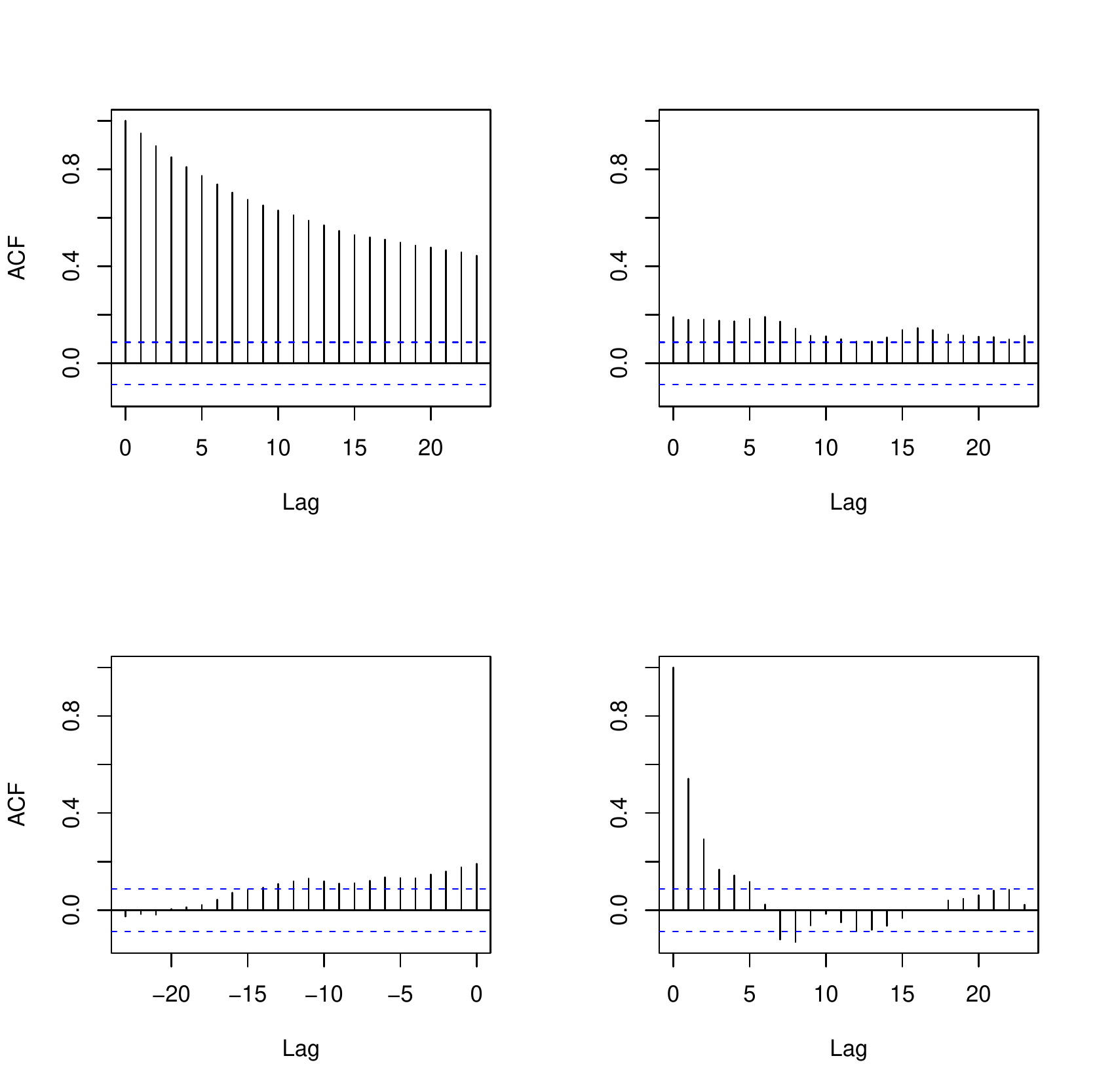}
  {\caption*{w=0.5}}
  \end{subfigure}
   \begin{subfigure}[t]{\textwidth}
   \centering
 \includegraphics[height=2.2in]{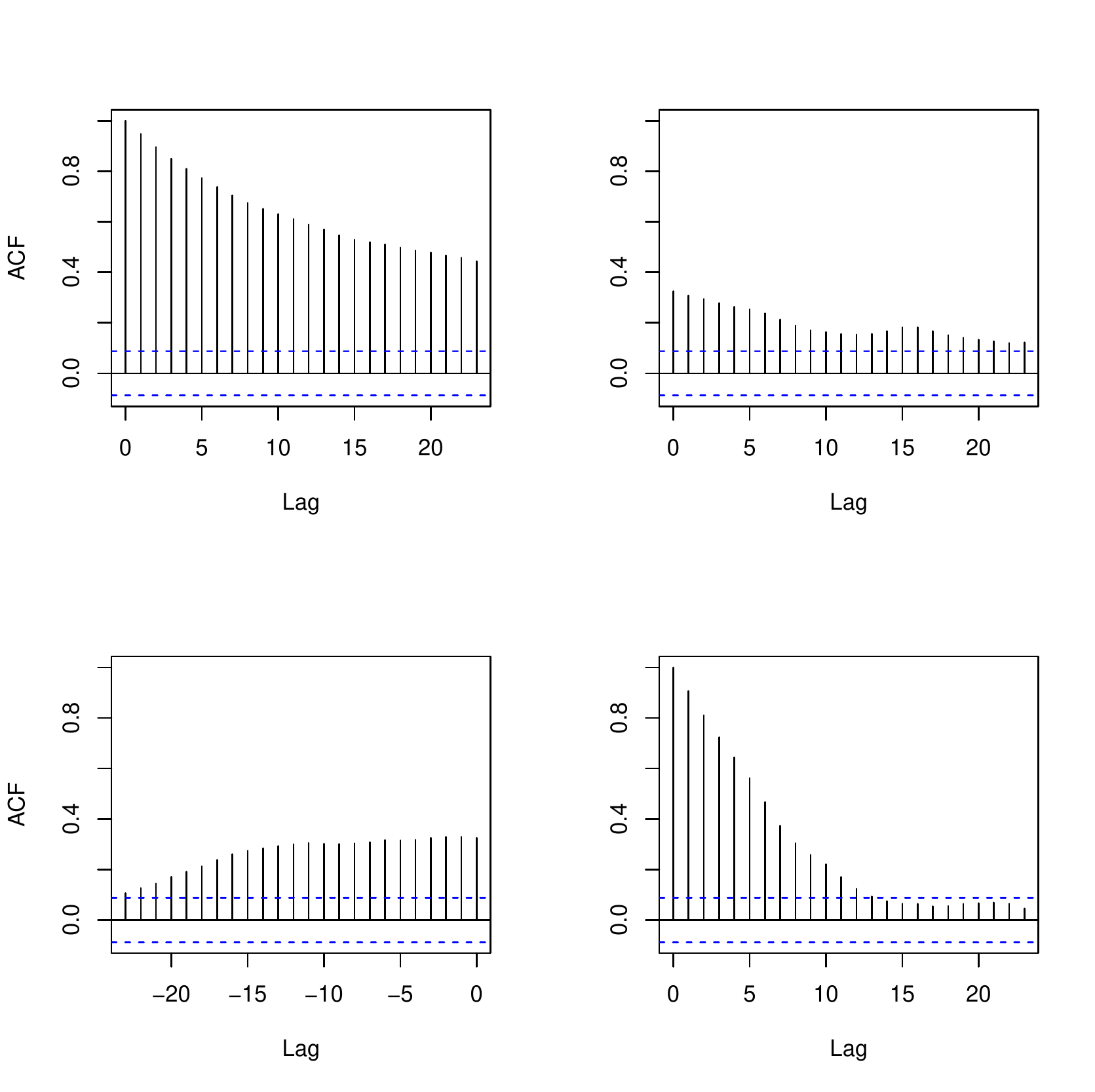}
 {\caption*{w=0.9}}
  \end{subfigure}
  \medskip
 Figure 15: Joint ACF of $R_{0,t}$ and $\hat{R}_{0,t}$
\end{figure}

\begin{center}
\begin{figure}
\centering
\includegraphics[width=5.0in,angle = 0]{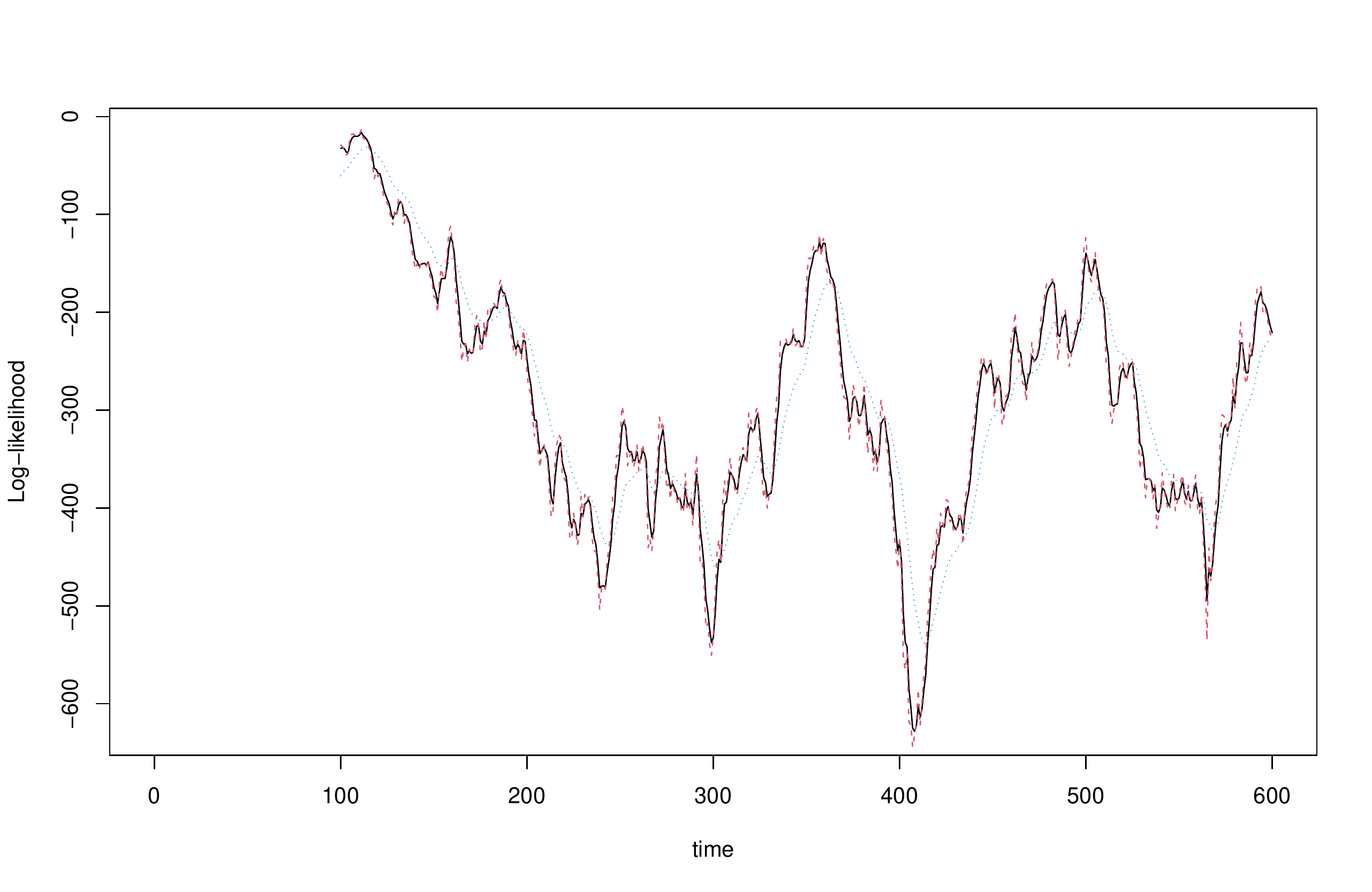}
{\caption*{Figure 16: Log-likelihoods, Stochastic $a$, \\
 red dashed line:w=0.1, black solid line: w=0.5, dotted green line: w=0.9 }}
\end{figure}
\end{center}

\end{document}